\documentclass[prx,aps,amssymb,twocolumn,noshowpacs]{revtex4-2}
\usepackage{color}
\usepackage{graphicx}
\usepackage{hyperref}
\usepackage{amsmath}
\usepackage{latexsym}
\usepackage{amssymb}
\usepackage{mathrsfs}
\usepackage{mathtools}
\usepackage{layout}
\usepackage{verbatim}
\usepackage{multirow}
\usepackage{bm}
\usepackage{amsfonts,epsfig}
\usepackage{lineno}

\begin{document} 
\title{Fast nuclear-spin gates and electrons-nuclei entanglement of neutral atoms in weak magnetic fields }

\date{\today}
\author{Xiao-Feng Shi}
\affiliation{School of Physics, Xidian University, Xi'an 710071, China}

\begin{abstract}
We present fast Rydberg-mediated entanglement involving nuclear spins of divalent atoms with $^{171}$Yb as an example. First, we show a nuclear-spin controlled phase gate of an arbitrary phase realizable either with two laser pulses when assisted by Stark shifts, or with three pulses. Second, we propose to create a state $(\lvert\text{cc}\rangle_{\text{e}} \otimes \lvert\Phi\rangle_{\text{n}}   + \lvert\Phi\rangle_{\text{e}}  \otimes \lvert\Psi\rangle_{\text{n}} )/\sqrt{2}$ entangled between the electrons~(e) and nuclear spins~(n) of two atoms, where $\lvert\Phi\rangle$ and $\lvert\Psi\rangle$ are two orthogonal Bell states and $\lvert \text{c}\rangle_{\text{e}}$ denotes the clock state. For want of a better term, it is called a Super Bell State for it mimics a ``large'' Bell state incorporating three ``smaller'' Bell states. Third, we show a protocol to create a three-atom state $(\sqrt{3}\lvert\text{ccc}\rangle_{\text{e}} \otimes \lvert\Lambda\rangle_{\text{n}}   + \lvert \text{W}\rangle_{\text{e}}  \otimes \lvert \text{GHZ}\rangle_{\text{n}} )/2$, where $\lvert\Lambda\rangle_{\text{n}}$ is a nuclear-spin state, $\lvert \text{W}\rangle_{\text{e}}$  is a W state in the ground-clock state space, and $\lvert \text{GHZ}\rangle_{\text{n}}$ is the Greenberger-Horne-Zeilinger~(GHZ) state in the nuclear-spin state space. The four protocols have a duration $x\pi/\Omega_{\text{m}}$ with $x$ about $2.6$~(two-pulse gate), $5.1$~(three-pulse gate), $7.7$, and $11$, respectively, where $\Omega_{\text{m}}$ is the largest Rabi frequency during the pulses. They possess high intrinsic fidelities, do not require single-site Rydberg addressing, and can be executed with large $\Omega_{\text{m}}$ in a weak, Gauss-scale magnetic field for they involve Rydberg excitation of both nuclear-spin qubit states in each atom. The latter two protocols can enable measurement-based preparation of Bell, hyperentangled, and GHZ states.

\end{abstract}
\maketitle

\section{introduction}\label{sec01}

\subsection{Fast nuclear-spin entanglement}\label{sec01A}
Dipole-dipole interactions of Rydberg atoms can provide efficient ways for  entanglement generation in individual alkali-metal atoms~\cite{PhysRevLett.85.2208,Lukin2001} as experimentally tested~\cite{Wilk2010,Isenhower2010,Zhang2010,Maller2015,Jau2015,Zeng2017,Levine2018,Picken2018,Levine2019,Graham2019,Jo2019,Fu2022,McDonnell2022,Bluvstein2022,Graham2022}. Meanwhile, Rydberg interactions in divalent alkaline-earth-metal or lanthanide atoms, which we call alkaline-earth-like~(AEL) atoms, such as ytterbium and strontium, can also create entanglement~\cite{Madjarov2020,Ma2022,Schine2022}. Compared to alkali-metal atoms, AEL atoms can be more easily cooled to low temperatures~\cite{Yamamoto2016,Saskin2018,Cooper2018,Norcia2018,Covey2019,Madjarov2020} and long-lived trapping of both the ground and Rydberg states is realizable~\cite{PhysRevLett.128.033201}. To date, the highest fidelity for entangling two neutral atoms was realized with Rydberg interactions of AEL atoms~\cite{Madjarov2020}. 

Another advantage in AEL is that for isotopes with nuclear spins, quantum information stored in nuclear spin states~\cite{Daley2008,Gorshkov2009} is insensitive to magnetic noise, can be preserved during laser cooling~\cite{Reichenbach2007}, and one neutral atom can host multiple, stable, and controllable nuclear-spin states useful for coding information~\cite{Gorshkov2009,Omanakuttan2021,Chen2022}. In these atoms, however, the g-factors of the ground and clock states are mainly of nuclear-spin character~\cite{Boyd2007} which can lead to simultaneous Rydberg excitation of both nuclear-spin qubit states in a Rydberg-mediated gate~\cite{Shi2021}. A useful protocol of entanglement generation is simultaneously exciting one out of the two qubit states in each atom to Rydberg states~\cite{Shi2019prap,Levine2019}, which means that the other qubit state in each atom shouldn't be Rydberg excited. 
To employ this method with nuclear-spin qubits, one can choose a Rydberg Rabi frequency $\Omega$ small compared to the Zeeman splitting $\Delta_{\text{Z}}$ in the Rydberg states as done in~\cite{Ma2022}, or use polarization rules~\cite{Shi2021}. Nevertheless, for nuclear spin qubits defined in the clock state, very large UV-laser Rydberg Rabi frequencies $\Omega$ can be realized~\cite{Madjarov2020,Chen2022} which seems an advantage, but the polarization method of~\cite{Shi2021} can't work, and using $\Omega\ll\Delta_{\text{Z}}$~(e.g., the experiment in~\cite{Ma2022} had $\Omega/\Delta_{\text{Z}}<0.1$) would either lengthen the gate duration with Gauss-scale B-fields, or bring extra magnetic noise when strong B-fields are used.

Here, we present nuclear-spin entangling gates realizable with a large $\Omega$ in a weak magnetic field when $\Omega/\Delta_{\text{Z}}\lessapprox1$. By exciting the four nuclear spin qubit states in two atoms simultaneously from the ground or clock state to Rydberg states, a two-qubit controlled phase gate of an arbitrary phase is realizable with three laser pulses of total duration about $5.05\pi/\Omega_{\text{m}}$, or with two pulses of total duration about $2.59\pi/\Omega_{\text{m}}$ when assisted by Stark shift, where $\Omega_{\text{m}}$ is the maximal Rabi frequency in the pulses. With $\Omega_{\text{m}}$ over $2\pi\times6$~MHz~\cite{Madjarov2020,Chen2022}, the gate duration can be less than $0.42$~($0.22)~\mu$s for the three~(two)-pulse gate, or even shorter for clock-Rydberg Rabi frequencies over $2\pi\times10$~MHz are realizable~\cite{Madjarov2020}. Most importantly, the gates are compatible with Gauss-scale magnetic fields as in recent nuclear-spin-qubit experiments~\cite{Barnes2022,Ma2022,Jenkins2022}.

\subsection{Supper Bell states}\label{sec01B}
We extend the idea in Sec.~\ref{sec01A} and present a three-pulse protocol to realize the following state entangled between the electrons~(e) and nuclear spins~(n) in two atoms via Rydberg blockade,
\begin{eqnarray}
 |\text{SBS}\rangle
 &\equiv&\frac{1}{\sqrt{2}}\bigg(\lvert\text{cc}\rangle_{\text{e}} \otimes \lvert\Phi\rangle_{\text{n}}   + \lvert\Phi\rangle_{\text{e}}  \otimes \lvert\Psi\rangle_{\text{n}} \bigg), \label{Bell}
\end{eqnarray}
where
\begin{eqnarray}
  \lvert\Phi\rangle_{\text{n}} &=& \frac{  \lvert\uparrow\downarrow\rangle_{\text{n}} + \lvert\downarrow\uparrow\rangle_{\text{n}} }{\sqrt{2}} , \nonumber\\
  \lvert\Psi\rangle_{\text{n}} &=&\frac{ e^{i\theta} \lvert\uparrow\uparrow\rangle_{\text{n}} + e^{i\theta'} \lvert\downarrow\downarrow\rangle_{\text{n}} }{\sqrt{2}}   \label{Bell2}
\end{eqnarray}
are two orthogonal Bell states entangled in the two nuclear spin states $\uparrow$ and $\downarrow$, $\theta$ and $\theta'$ are two angles, and 
\begin{eqnarray}
  \lvert\Phi\rangle_{\text{e}} &=&  \frac{ \lvert\text{cg}\rangle_{\text{e}} +\lvert\text{gc}\rangle_{\text{e}}    }{\sqrt{2}} \nonumber\label{Bell3}
\end{eqnarray}
is a Bell state entangled in the electronic ground~(g) and clock~(c) states. Because of the similar structure, $ \lvert\Phi\rangle_{\text{n}}$ and $ \lvert\Phi\rangle_{\text{e}}$ are labeled by the same Greek letter. Looking at it as a whole, Eq.~(\ref{Bell}) is entangled between the two-atom electronic states~($\lvert\text{cc}\rangle_{\text{e}} , \lvert\Phi\rangle_{\text{e}} $) and two-atom nuclear spin states~($\lvert\Phi\rangle_{\text{n}}, \lvert\Psi\rangle_{\text{n}}$). For want of better term, Eq.~(\ref{Bell}) is called a supper Bell state~(SBS) because it is like a large Bell state including three ``small'' Bell states. To our knowledge, no such exotic two-particle entangled state containing three Bell states was reported. 

SBS is prepared by three laser pulses, two UV laser pulses for the clock-Rydberg transition and one laser pulse for the ground-Rydberg transition. The three pulses have a total duration $3.4\pi/\Omega_{\text{eff}}$~(or $7.7\pi/\Omega$), where $\Omega_{\text{eff}}$ is the Rabi frequency for the ground-Rydberg transition and $\Omega$ is the largest UV laser Rabi frequency among the two UV laser pulses for the clock-Rydberg transition.

\subsection{A three-atom state including W and GHZ states}\label{sec01C}
Our theory can be used to realize the following state
\begin{eqnarray}
 \lvert  \blacktriangle \rangle&=&\frac{1}{2}\Big[(\sqrt{3}\lvert\text{ccc}\rangle_{\text{e}} \otimes \lvert\Lambda\rangle_{\text{n}}   + \lvert \text{W}\rangle_{\text{e}}  \otimes \lvert \text{GHZ}\rangle_{\text{n}} )  \Big],  \label{W01}
\end{eqnarray}
which has rich entanglement in three atoms, where
\begin{eqnarray}
 \lvert\Lambda\rangle_{\text{n}}&=&\frac{1}{\sqrt{6}}\Big[ \lvert \uparrow \uparrow \downarrow\rangle_{\text{n}} +  \lvert \uparrow\downarrow\uparrow \rangle_{\text{n}} +  \lvert \downarrow\uparrow\uparrow\rangle_{\text{n}} + e^{i\Theta}( \lvert \uparrow\downarrow\downarrow \rangle_{\text{n}} +  \lvert \downarrow\uparrow\downarrow \rangle_{\text{n}} \nonumber\\
 &&+  \lvert \downarrow\downarrow\uparrow \rangle_{\text{n}})  \Big] \nonumber
\end{eqnarray}
is the sum of two different nuclear-spin W states with a relative phase, 
\begin{eqnarray}
\lvert \text{W}\rangle_{\text{e}} &=&  \frac{1}{\sqrt{3}}\left(\lvert\text{gcc}\rangle_{\text{e}}+  \lvert\text{cgc}\rangle_{\text{e}}+ \lvert\text{ccg}\rangle_{\text{e}}  \right) \nonumber
\end{eqnarray}
is an electronic W state~\cite{Dur2000,Fang2019} which is maximally entangled in the ground-clock state space, and 
\begin{eqnarray}
\lvert \text{GHZ}\rangle_{\text{e}} &=&  \frac{1}{\sqrt{2}}\left( \lvert \uparrow  \uparrow \uparrow \rangle_{\text{n}} + \lvert \downarrow\downarrow \downarrow \rangle_{\text{n}} \right)\nonumber
\end{eqnarray}
is a Greenberger-Horne-Zeilinger~(GHZ) state~\cite{GHZ1989} which is maximally entangled in the nuclear-spin state space. 

Like SBS, the state in Eq.~(\ref{W01}) is prepared by two UV laser pulses for the clock-Rydberg transition and one laser pulse for the ground-Rydberg transition with a total duration $3.5\pi/\Omega_{\text{eff}}$~(or $11\pi/\Omega$). Because $\Omega_{\text{eff}}$ is in general small~\cite{Shi2021}, the speed for creating SBS and $\lvert  \blacktriangle \rangle$ is bottlenecked by the available laser powers for realizing the ground-Rydberg transition.

The remainder of this paper is organized as follows. In Sec.~\ref{Sec02}, we study a three-pulse protocol to realize a quantum gate in the nuclear spins of the clock or ground states. In Sec.~\ref{Sec03}, we study a similar nuclear-spin quantum gate realized by two laser pulses when assisted by Stark shift. In Sec.~\ref{Sec04}, we show a three-pulse protocol to create SBS. In Sec.~\ref{Sec05}, we show a three-pulse protocol to create a three-atom state which contains W and GHZ states simultaneously. In Sec.~\ref{Sec06}, we discuss creation of Bell, hyperentangled, and GHZ states by measuring the states studied in Secs.~\ref{Sec04}~ and~\ref{Sec05}. Sections~\ref{Sec07} and~\ref{Sec08} give discussions and conclusions, respectively.

\section{Fast Nuclear-spin Quantum gates}\label{Sec02}
\subsection{A controlled-phase gate of any desired phase}\label{Sec02A}
We first show a sequence to realize a nuclear-spin quantum gate of the form,
\begin{eqnarray}
 &&\lvert \text{c}_\uparrow \text{c}_\uparrow\rangle \rightarrowtail{ } e^{-i\alpha}\lvert \text{c}_\uparrow \text{c}_\uparrow\rangle,\nonumber\\
&& \lvert \text{c}_\uparrow \text{c}_\downarrow\rangle \rightarrowtail e^{-i\beta/2} \lvert \text{c}_\uparrow \text{c}_\downarrow\rangle ,\nonumber\\
&& \lvert \text{c}_\downarrow \text{c}_\uparrow\rangle \rightarrowtail e^{-i\beta/2} \lvert \text{c}_\downarrow \text{c}_\uparrow\rangle,\nonumber\\
&& \lvert \text{c}_\downarrow \text{c}_\downarrow\rangle \rightarrowtail e^{i\alpha } \lvert \text{c}_\downarrow \text{c}_\downarrow\rangle , \label{Gate01} 
\end{eqnarray}
where the first~(second) $\text{c}_{\uparrow(\downarrow)}$ represents the state of the first~(second) atom, $\alpha$ and $\beta$ are angles where $\beta$ is determined by a global laser phase, and 
\begin{eqnarray}
\lvert \text{c}_\uparrow\rangle&\equiv& | \text{c}\rangle_{\text{e}}\otimes \lvert\uparrow\rangle_{\text{n}},\nonumber\\
\lvert \text{c}_\downarrow\rangle&\equiv& | \text{c}\rangle_{\text{e}}\otimes \lvert\downarrow\rangle_{\text{n}},\nonumber\\
\lvert \text{g}_\uparrow\rangle&\equiv& |\text{g}\rangle_{\text{e}}\otimes \lvert\uparrow\rangle_{\text{n}},\nonumber\\
\lvert \text{g}_\downarrow\rangle&\equiv& |\text{g}\rangle_{\text{e}}\otimes \lvert\downarrow\rangle_{\text{n}},\nonumber
\end{eqnarray}
where e~(n) denotes the electronic~(nuclear spin) state of the atom, $\otimes$ is used because for good approximation, the electron and nuclear spin are decoupled in the ground~(g) and clock~(c) states ~\cite{Shi2021,Shi2021pra,Chen2022}. The gate in Eq.~(\ref{Gate01}) is a controlled-phase gate because by using the single-qubit phase gates 
\begin{eqnarray}
\lvert \text{c}_\uparrow\rangle&\rightarrowtail& e^{i\alpha/2} \lvert \text{c}_\uparrow\rangle,\nonumber\\
\lvert \text{c}_\downarrow\rangle&\rightarrowtail& e^{i(\beta-\alpha)/2} \lvert \text{c}_\downarrow\rangle,\nonumber
\end{eqnarray}
in both atoms~\cite{Shi2021,Shi2021pra}, the gate in Eq.~(\ref{Gate01}) becomes 
\begin{eqnarray}
 &&\lvert \text{c}_\uparrow \text{c}_\uparrow\rangle \rightarrowtail \lvert \text{c}_\uparrow \text{c}_\uparrow\rangle ,\nonumber\\
&& \lvert \text{c}_\uparrow \text{c}_\downarrow\rangle \rightarrowtail  \lvert \text{c}_\uparrow \text{c}_\downarrow\rangle ,\nonumber\\
&& \lvert \text{c}_\downarrow \text{c}_\uparrow\rangle \rightarrowtail  \lvert \text{c}_\downarrow \text{c}_\uparrow\rangle,\nonumber\\
&& \lvert \text{c}_\downarrow \text{c}_\downarrow\rangle \rightarrowtail  e^{i\beta} \lvert \text{c}_\downarrow \text{c}_\downarrow\rangle,\label{Gate02} 
\end{eqnarray}
where $\beta$ is adjustable via laser phases. The case $\beta=\pi$ corresponds to the canonical CZ gate as realized in the Rydberg quantum gate~\cite{Saffman2010}. In principle, protocols used in alkali-metal atoms~\cite{Shi2021qst} can also be used for creating the gate in Eq.~(\ref{Gate02}). But for nuclear spins in AEL, two nearby nuclear spin states in either the ground or clock states are nearly degenerate in a weak B-field, so that both nuclear spin qubit states can be excited to Rydberg states~\cite{Shi2021}. To use protocols tested with alkali-metal atomic hyperfine qubits for entangling AEL nuclear-spin qubits, one may use strong magnetic fields to suppress the Rydberg excitation of the nontarget nuclear spin states~\cite{Chen2022}. The benefit of using strong magnetic fields is that the polarization of the laser fields can fluctuate without decreasing the gate fidelity too much~\cite{Chen2022}. On the other hand, strong magnetic fields can lead to large field fluctuation in an array of atoms. A compromise is to use quite small Rydberg Rabi frequency when a weak magnetic field is employed, as in the experiment reported in Ref.~\cite{Ma2022}. This raises a question whether fast nuclear-spin quantum gates can be created in a weak magnetic field. Below, we first show the physical possibility for our gate protocols and then show a gate realizable with large Rydberg Rabi frequencies in a weak magnetic field.

\begin{figure}
\includegraphics[width=2.0in]
{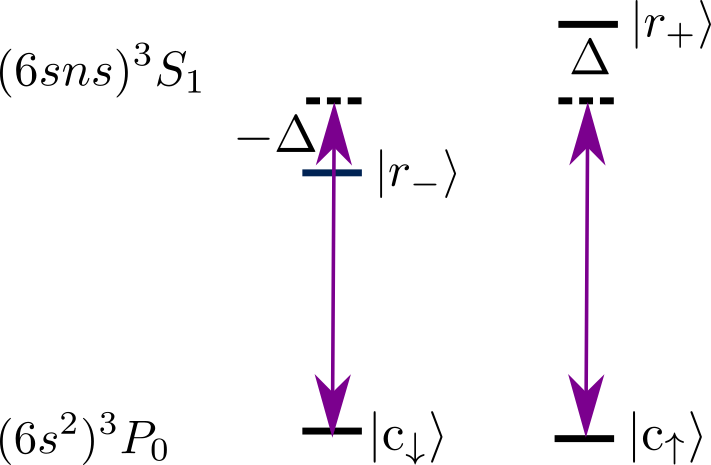}
\caption{ Rydberg excitation of two nuclear-spin qubit states $\lvert\uparrow\rangle$ and $\lvert\downarrow\rangle$ in $\lvert \text{c}_\uparrow\rangle\equiv\lvert (6s^2)^3P_0 \lvert I=1/2, m_I=1/2\rangle$ and $\lvert\text{c}_\downarrow\rangle\equiv \lvert (6s^2)^3P_0 \lvert I=1/2, m_I=-1/2\rangle$ to two respective Rydberg states $\lvert r_-\rangle$ and $\lvert r_+\rangle$, where c denotes the clock state with $^{171}$Yb as an example. In a B-field of several Gauss~(1~G=$10^{-4}$~T), the two nuclear spin states in the clock state are nearly degenerate compared to the MHz-scale Rabi frequencies considered in this paper. The laser fields are $\pi$ polarized and tuned to the middle of the gap between $\lvert r_+\rangle$ and $\lvert r_-\rangle$. \label{figure1} }
\end{figure}
\subsection{Laser excitation of the ground-Rydberg and clock-Rydberg transitions}\label{Sec02B}
The gate protocols in this paper depend on the possibility to excite the ground and clock states to Rydberg states with MHz-scale Rabi frequencies, for which we analyze in detail below. 

The ground state can be excited to a $(6s6n)~^3S_1$ Rydberg state via the largely detuned $(6s6p)~^3P_1$ state as theoretically analyzed~\cite{Shi2021} and experimentally verified~\cite{Burgers2022,Ma2022}. The hyperfine interaction can mix the singlet and triplet states~\cite{Ding2018}, but for the $F=I+1=3/2$ state there is no mixing~\cite{Lurio1962}. In this paper, we consider exciting the ground or clock states to the $F=I+1$ manifold of $(6s6n)~^3S_1$ state with $n\sim70$, for which the nearby Rydberg states~(of different $F$) are separated by more than 1~GHz~\cite{Shi2021}, which is orders of magnitude larger than the Rydberg Rabi frequency $\Omega$ considered in this paper. So, we can ignore the excitation of the nearby Rydberg states. Reference~\cite{Shi2021} showed that a ground-Rydberg Rabi frequency $2\pi\times1.4$~MHz can be realized for a Rydberg state of principal quantum $70$. As for the clock state, the experiment of Ref.~\cite{Madjarov2020} excited the clock state of strontium to a Rydberg state of $n=61$ with a Rabi frequency up to $2\pi\times 13$~MHz, and large Rabi frequencies were adopted in theoretical analyses~\cite{Chen2022}.

We consider a Gauss-scale magnetic field for specifying the quantization axis, so that the two Zeeman substates $\lvert\uparrow\rangle\equiv|m_I=1/2\rangle$ and $\lvert\downarrow\rangle\equiv|m_I=-1/2\rangle$ in the ground~(or clock)~state can be assumed degenerate~\cite{Shi2021}. In the $F=3/2$ level of the $(6s6n)~^3S_1$ Rydberg state, there is a frequency separation $2\Delta\approx2\pi\times1.9B$~MHz$/$G~\cite{Chen2022} between the two hyperfine substates $|r_\pm\rangle\equiv (6s6n)~^3S_1\lvert F=3/2,m_F=\pm1/2\rangle$ , where $B$ is the magnetic field in units of Gauss.

According to the angular momentum selection rule, one can find that the Rabi frequencies $ \Omega_{\text{g}\uparrow(\downarrow)}$ for the ground-Rydberg transition via the $(6s6p)~^3P_1$ state, and the Rabi frequencies $ \Omega_{\text{cr}\uparrow(\downarrow)}$ for the clock-Rydberg transition satisfy the condition~\cite{Shi2021}
\begin{eqnarray}
\Omega_{\text{g}\uparrow}&=&-\Omega_{\text{g}\downarrow},\nonumber\\ 
\Omega_{\text{c}\uparrow}&=&\Omega_{\text{c}\downarrow},\label{RabiRelation}
\end{eqnarray}
where the ground-Rydberg transition will be used in the creation of SBS and $\lvert \blacktriangle\rangle$. Note that Eq.~(\ref{RabiRelation}) does not mean that the four entanglement protocols are limited to the forms shown. The nuclear-spin quantum gates in Secs.~\ref{Sec02C} and~\ref{Sec03} can also be executed by ground-Rydberg transitions for nuclear spins in the ground state, and other forms of SBS and $\lvert \blacktriangle\rangle$ can be created with the ``g'' and ``c''
 states exchanged in the equations defining them.

For the nuclear-spin gates and the first two pulses of the entanglement protocols in Secs.~\ref{Sec04} and~\ref{Sec05}, the laser is tuned to the middle of the gap between the two Rydberg states $\lvert r_+\rangle$ and $\lvert r_-\rangle$ as shown in Fig.~\ref{figure1}, i.e., the detuning of the Rydberg lasers are $\Delta$ and $-\Delta$ for $\lvert r_+\rangle$ and $\lvert r_-\rangle$, respectively, where the detuning is given by the transition frequency deducted by the laser frequency. The magnetic field is fixed, while the laser fields can be tuned so that Rydberg Rabi frequencies can be different in different pulses.

\begin{figure}
\includegraphics[width=3.2in]
{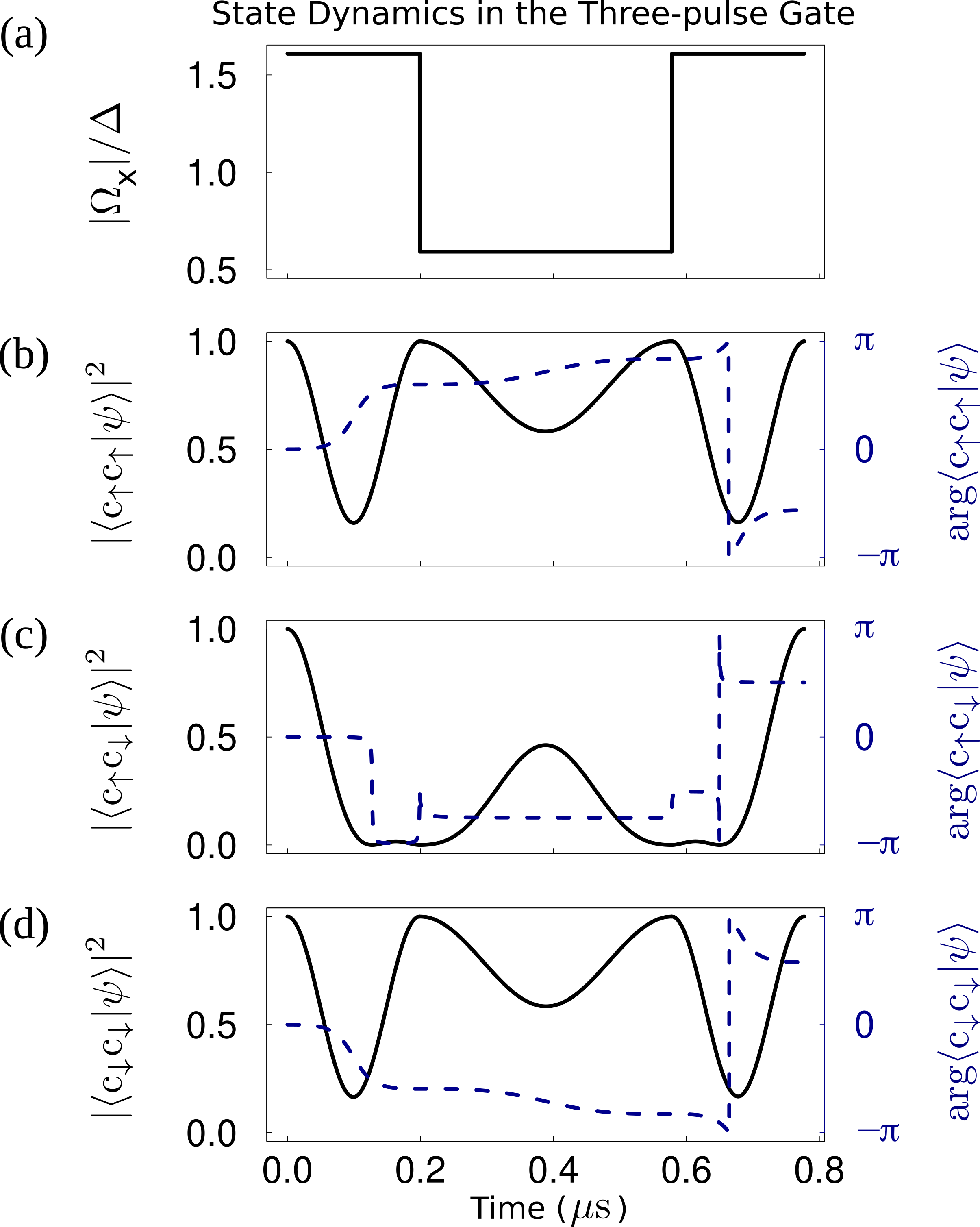}
 \caption{State dynamics for different input states when ignoring Rydberg-state decay in the controlled-phase gate~(with $\beta=-\pi$ as an example here). (a) $| \Omega_{\text{x}} |/\Delta$ as a function of time during the three pulses of the nuclear-spin gate. Here, the largest Rabi frequency is $| \Omega_{\text{1}} |=| \Omega_{\text{3}}| = 2\pi \times 3.25$~MHz, and the relative phases of the three Rabi frequencies are $0,3\pi/4, -\pi/2$. The Rydberg interaction is $V_0/2\pi=260$~MHz~(its fluctuation is studied in Sec.~\ref{Sec02D}). (b), (c), and (d) show the population and phase of the ground-state component in the wavefunction when the input states for the gate protocol are $\lvert \text{c}_\uparrow \text{c}_\uparrow\rangle$, $\lvert \text{c}_\uparrow \text{c}_\downarrow\rangle$, and $\lvert \text{c}_\downarrow \text{c}_\downarrow\rangle$, respectively. The final population errors in the target ground states are $2.6\times10^{-5}$, $2.1\times10^{-5}$, and $2.4\times10^{-5}$ in (b), (c), and (d), respectively, and their final phases are $-1.771$, $1.587$, and $1.816$~rad, respectively. The state dynamics for the input state $\lvert \text{c}_\downarrow \text{c}_\uparrow\rangle$ is similar to that of $\lvert \text{c}_\uparrow \text{c}_\downarrow\rangle$.      }\label{Gate-unitary}
\end{figure}

\subsection{A three-pulse sequence}\label{Sec02C}
We consider a controlled-phase gate with three sequential laser pulses sent to the two atoms. The Hamiltonians for the gate are shown in Appendix~\ref{AppendixA}. According to Refs.~\cite{Shi2017,Shi2018prapp2} and the Rydberg blockade condition~\cite{PhysRevLett.85.2208}, by applying a pulse of duration $\mathbb{T}_1=2\pi/\sqrt{\Delta^2+2\Omega_{\text{x}}^2}$, where $\Omega_{\text{c}\uparrow}=\Omega_{\text{c}\downarrow}\equiv\Omega_{\text{x}}$~[according to Eq.~(\ref{RabiRelation})] is the Rabi frequency in the x-th pulse, $x=1,2$, or $3$, we have the state map
\begin{eqnarray}
 &&\lvert \text{c}_\uparrow \text{c}_\uparrow\rangle \xrightarrow{ } e^{-i(1+\Delta/\sqrt{\Delta^2+2\Omega_{\text{x}}^2}) }\lvert \text{c}_\uparrow \text{c}_\uparrow\rangle,\nonumber\\
&& \lvert \text{c}_\downarrow \text{c}_\downarrow\rangle \xrightarrow{ } e^{-i(1-\Delta/\sqrt{\Delta^2+2\Omega_{\text{x}}^2}) }\lvert \text{c}_\downarrow \text{c}_\downarrow\rangle. \nonumber\label{gate-pulse1}
\end{eqnarray}
For the first pulse, we find that when 
\begin{eqnarray}
 \Omega_{\text{1}} /\Delta=  1.6088, \nonumber
\end{eqnarray}
the states $\lvert \text{c}_\uparrow \text{c}_\downarrow\rangle$ and $\lvert \text{c}_\downarrow \text{c}_\uparrow\rangle$ are excited to Rydberg states. For example, driven by the Hamiltonian $(\Omega_{\text{c}\uparrow}\lvert r_+\rangle\langle\text{c}_\uparrow\rvert  +\Omega_{\text{c}\downarrow}\lvert r_-\rangle\langle\text{c}_\downarrow\rvert+\text{H.c.})/2 +\Delta(\lvert r_+\rangle\langle r_+\rvert-\lvert r_-\rangle\langle r_-\rvert)$ for each atom, $\lvert \text{c}_\uparrow \text{c}_\downarrow\rangle$ evolves to 
\begin{eqnarray}
(e^{i\varphi_+}\lvert r_+\text{c}_\downarrow\rangle + e^{i\varphi_-}\lvert  \text{c}_\uparrow r_-\rangle)/\sqrt{2},\label{Gate-p2}
\end{eqnarray}
where $\varphi_+=2.645$~rad and $\varphi_-=0.497$~rad. To map the state down to ground states, we find that it is necessary to change the phase of the Rydberg state, which is why we apply a second pulse in the condition 
\begin{eqnarray}
 \Omega_{\text{2}} /\Delta=  0.5932e^{i3\pi/4}.\nonumber
\end{eqnarray}
After the second pulse, the state in Eq.~(\ref{Gate-p2}) becomes $e^{-i\varphi_+}\lvert r_+\text{c}_\downarrow\rangle + e^{-i\varphi_-}\lvert  \text{c}_\uparrow r_-\rangle)/\sqrt{2}$. Then, with a third pulse of condition
\begin{eqnarray}
 \Omega_{\text{3}} &=& \Omega_{\text{1}} e^{i\beta/2},\nonumber 
\end{eqnarray}
 a gate in Eq.~(\ref{Gate01}) is realized with $\alpha=1.793$~rad in the ideal blockade condition. If $\beta=\pm\pi$ is chosen, a CZ gate is realized, for which a numerical simulation is shown in Fig.~\ref{Gate-unitary} with a finite $V$. The total duration of the gate is $
 \sum_{x=1}^3 \mathbb{T}_x \approx \frac{5.054\pi}{ \Omega_{\text{1}} }$, which is in terms of the maximal Rabi frequency $\Omega_{\text{1}}$~(or $\Omega_{\text{3}}$) during the three pulses. If we assume $| \Omega_{\text{1}} |/2\pi = 3.25$~MHz~(corresponding to $\Delta/2\pi=2.02$~MHz in a B-field of $2.1$~G), the gate duration would be $0.78~\mu$s. We use $| \Omega_{\text{x}} |\leq2\pi\times3.25$~MHz in the numerical example for it is equal to the maximal Rabi frequencies used later on in creating SBS and $\lvert \blacktriangle\rangle$ in Secs.~\ref{Sec03} and~\ref{Sec04}, but in experiments much larger UV Rydberg Rabi frequencies are realizable since strong UV fields are available; for example, Rydberg Rabi frequencies up to $2\pi\times13$~MHz were realized for exciting a Rydberg state of $n=61$ with optimization of laser system and beam path~\cite{Madjarov2020}.

\subsection{Numerical analyses}\label{Sec02D}
Here, we numerically study the fidelity of the nuclear-spin gate. Because a CZ gate has the maximal entangling power among the two-qubit gates~\cite{Williams2011}, we study the case with $\beta=-\pi$. Three main factors limit the intrinsic fidelity, the Rydberg state decay, the finiteness of $V$, and the fluctuation of $V$. The error due to Rydberg-state decay is~\cite{Saffman2005}
\begin{eqnarray}
 E_{\text{decay}} &=&t_{\text{Ryd}}/\tau,\label{decayE}
\end{eqnarray}
where $t_{\text{Ryd}}$ is the Rydberg superposition time and $\tau$ is the lifetime of the Rydberg state. We numerically found $
t_{\text{Ryd}} \approx  \frac{2.7\pi}{  \Omega_{\text{1}}}$. If the maximal Rabi frequency $| \Omega_{\text{1}} |$ is $2\pi\times3.25$~MHz, the decay-induced error is $E_{\text{decay}}=1.26\times10^{-3}$ with $\tau=330~\mu$s~\cite{Shi2021}.

The analyses in Sec.~\ref{Sec02C} assumed infinite Rydberg interactions, but a blockade error will arise with finite $V$~\cite{Saffman2005,Shi2018prapp2}. The experiment of Ref.~\cite{Ma2022} estimated that the $C_6$ coefficient for the $\lvert (6sns)^3S_1, F=I+1\rangle$ state with $n=50$ can be up to $2\pi\times15$~GHz~$\mu m^6$. With the $n^{11}$ scaling of van der Waals interaction, this would imply $C_6/2\pi=15\cdot(70/50)^{11}\approx607$~GHz~$\mu m^6$ for the state we consider. To have a more conserved estimate about the gate fidelity in this paper, we use a smaller $C_6/2\pi=192$~GHz~$\mu m^6$ from the analyses of Ref.~\cite{Shi2021}. We consider an atomic separation of $3~\mu$m~(this short atomic separation is possible since a value $2.4~\mu$m was realized in the experiment of Ref.~\cite{Ma2022}), for which we have $V_0/2\pi\approx260$~MHz. The gate protocols in this paper do not need single-site addressing since all the pulses simultaneously excite all atoms. So, the small atomic separation does not bring extra difficulty. There is nearly no population loss to the states caused by the finiteness of $V$ as shown in Fig.~\ref{Gate-unitary}. With the finite $V$, the final phases for the four different input states differ from the desired ones. We numerically found that the protocol maps the states $ \{\lvert \text{c}_\uparrow \text{c}_\uparrow\rangle,\lvert \text{c}_\uparrow \text{c}_\downarrow\rangle, \lvert \text{c}_\downarrow \text{c}_\uparrow\rangle, \lvert \text{c}_\downarrow \text{c}_\downarrow\rangle  \}$ to $ \{e^{-i\alpha'}\lvert \text{c}_\uparrow \text{c}_\uparrow\rangle,e^{-i\beta'/2}\lvert \text{c}_\uparrow \text{c}_\downarrow\rangle$$, e^{-i\beta'/2}\lvert \text{c}_\downarrow \text{c}_\uparrow\rangle$$, e^{i\alpha''}\lvert \text{c}_\downarrow \text{c}_\downarrow\rangle  \}$ when the interaction is $V=V_0$, where $ \{ \alpha' , \beta'/2,\alpha''\} \approx \{1.771,-1.587,1.816\}$~rad. This means that by using $
\lvert \text{c}_\uparrow\rangle\rightarrowtail e^{i\alpha'/2} \lvert \text{c}_\uparrow\rangle,
\lvert \text{c}_\downarrow\rangle\rightarrowtail e^{i(-\alpha'+\beta')/2} \lvert \text{c}_\downarrow\rangle$, a gate of map diag$\{1,1,1, e^{i(\alpha''-\alpha'+\beta')} \}$ is realized, where $\alpha''-\alpha'+\beta'\approx-3.128 $~rad, which differs from $-\pi$ by about $0.0135$~rad. To quantify the gate fidelity, we define
\begin{eqnarray}
\hat{U}= \text{diag} \{e^{i(\theta-\alpha')} ,e^{-i(\beta'/2+\theta)} , e^{-i(\beta'/2+\theta)}  , e^{i(\alpha''+\theta)} \}\nonumber\\\label{GateV}
\end{eqnarray}
as the target gate map, where $\theta = -(\pi+\alpha''-\alpha'+\beta' ) /4$. Then, using the single-qubit phase gates 
\begin{eqnarray}
\lvert \text{c}_\uparrow\rangle&\rightarrowtail& e^{i(\alpha'-\theta)/2} \lvert \text{c}_\uparrow\rangle,\nonumber\\
\lvert \text{c}_\downarrow\rangle&\rightarrowtail& e^{-i(\alpha'-\theta)/2+i(\beta'/2+\theta)} \lvert \text{c}_\downarrow\rangle,\nonumber
\end{eqnarray}
in both atoms one can transform Eq.~(\ref{GateV}) to the canonical CZ gate.

\begin{figure}
\includegraphics[width=2.5in]
{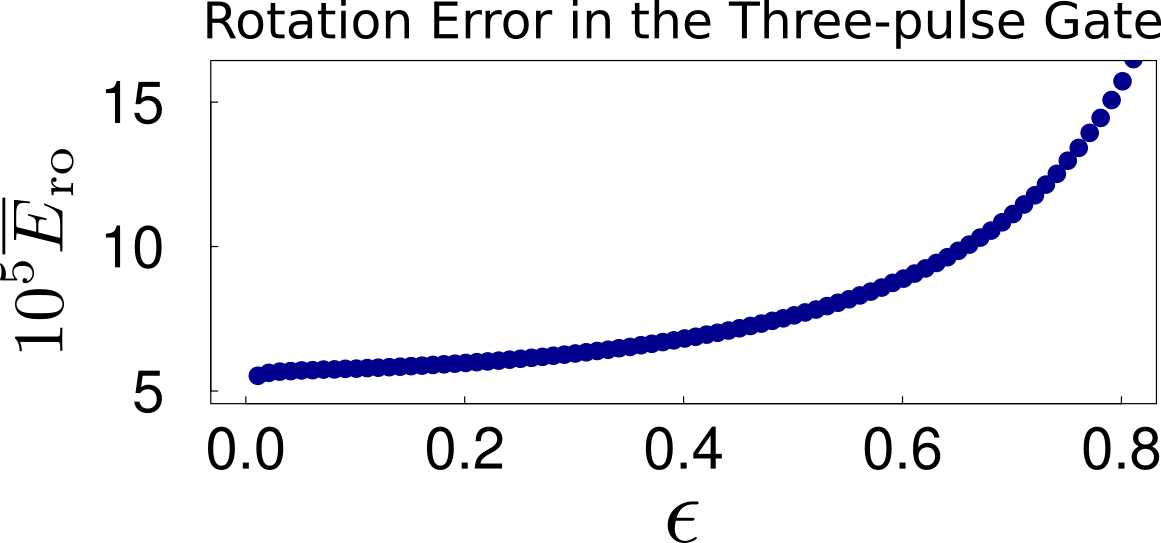}
 \caption{ Rotation error of the nuclear-spin gate~(rescaled by $10^5$) of Eq.~(\ref{GateF01}) averaged by uniformly varying the Rydberg interaction $V$ in $[(1-\epsilon)V_0,~(1+\epsilon)V_0]$, where $V_0/2\pi=260$~MHz. The fluctuation of $V$ leads to phases different from the desired phases in Eq.~(\ref{GateV}). The largest Rabi frequency is $| \Omega_{\text{1}} |/2\pi = 3.25$~MHz shown in Fig.~\ref{Gate-unitary}(a). The gate fidelity $1- \overline{E}_{\text{ro}} -  E_{\text{decay}} $ is over $0.998$ for the $\epsilon$ shown here, where the gate error is dominated by the Rydberg-state decay.   }\label{Gate-rotation}
\end{figure}

Beside that the finiteness of $V$ can cause error, the fluctuation of $V$ results in error, too. We define the rotation error by~\cite{Pedersen2007}
\begin{eqnarray}
 E_{\text{ro}} &=& 1-\frac{1}{20}\left[|\text{Tr}(\hat{U}^\dag \hat{\mathscr{U}})|^2 + \text{Tr}(\hat{U}^\dag\hat{\mathscr{U}}\hat{\mathscr{U}}^\dag\hat{U} )\right] ,\label{GateF01}
\end{eqnarray}
where $\hat{\mathscr{U}}=$diag$\{e^{i\eta_1}, e^{i\eta_2}, e^{i\eta_3}, e^{i\eta_4} \}$  is the state transform matrix with a fluctuating $V$, i.e., a $V$ not equal to $V_0$. In order to evaluate the effect of fluctuation, we consider the average 
\begin{eqnarray}
 \overline{E}_{\text{ro}} &=& \frac{ \int E_{\text{ro}}(V) dV}{\int dV}, \label{RoErrorV}
\end{eqnarray}
where the integration is over $V\in[(1-\epsilon)V_0,~(1+\epsilon)V_0]$; we use a uniform distribution for it can lead to a larger gate error compared to a Gaussian distribution so that we can estimate the lower bound for the gate fidelity. The results shown in Fig.~\ref{Gate-rotation} indicate that the fluctuation of $V$ does not result in small fidelities. Even when $V$ tends to deviate from the desired interaction with $\epsilon=0.8$, we numerically found that the fidelity is $1- \overline{E}_{\text{ro}} -  E_{\text{decay}} \approx0.9986$. 

\begin{figure}
\includegraphics[width=3.2in]
{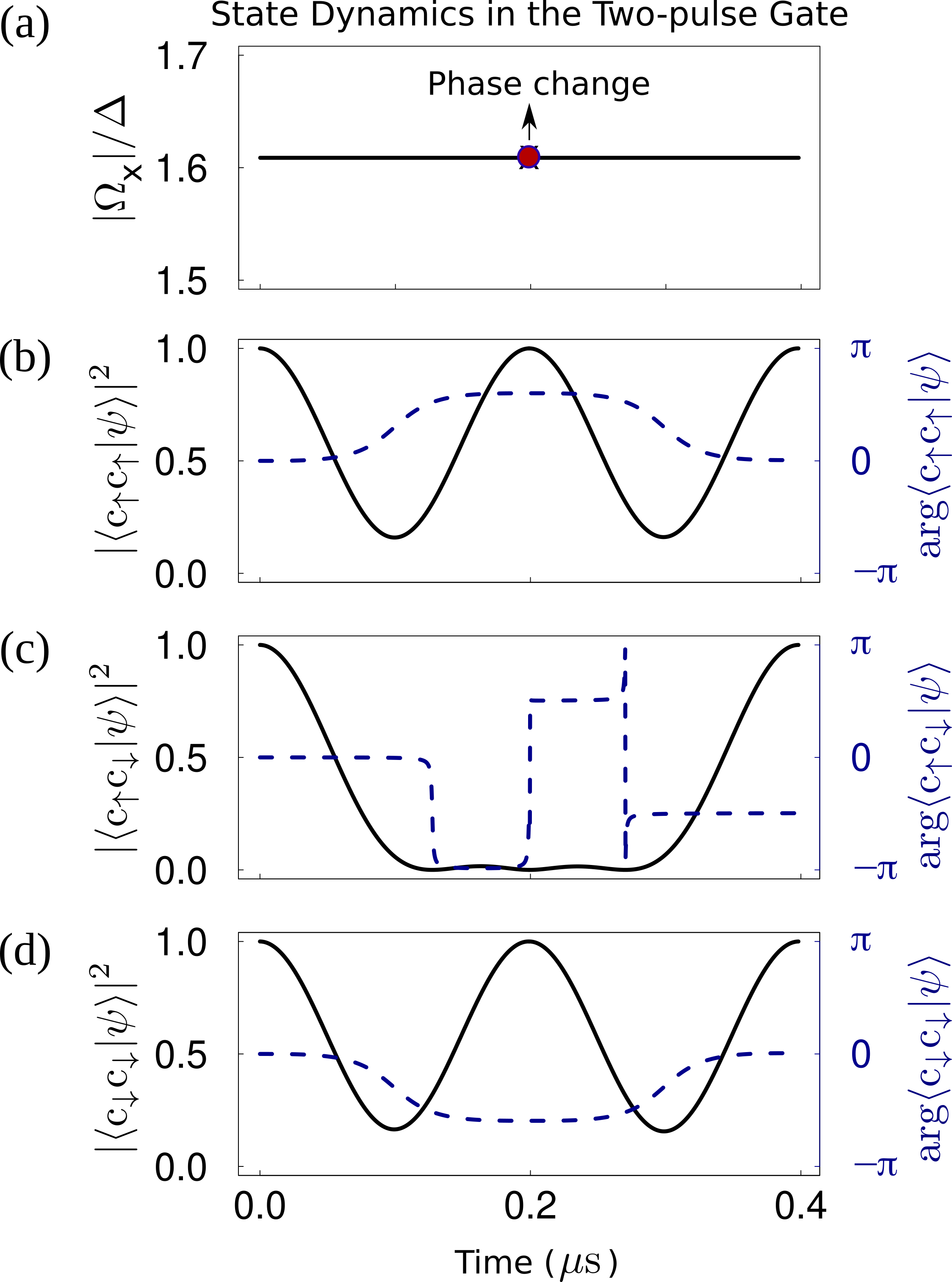}
 \caption{State dynamics of the two-pulse nuclear-spin gate with $\beta=\pi$ and $|\Omega_{\text{2}}|= \Omega_{\text{1}}  \approx 2\pi\times3.25$~MHz as an example. (a) The two laser pulses have the same strength and duration. The relative phase of the two Rabi frequencies are $0$ and $\pi/2+\kappa$, where $\kappa$ is given above Eq.~(\ref{kappaEquation}). A Stark shift $4\Delta\approx8.08$~MHz was assumed in $\lvert r_-\rangle$ during the second pulse, and the frequency of the laser during the second pulse increases by $2\Delta$ compared to the first pulse. The Rydberg interaction is $V_0/2\pi=260$~MHz. (b), (c), and (d) show the population and phase of the ground-state component in the wavefunction when the input states for the gate protocol are $\lvert \text{c}_\uparrow \text{c}_\uparrow\rangle$, $\lvert \text{c}_\uparrow \text{c}_\downarrow\rangle$, and $\lvert \text{c}_\downarrow \text{c}_\downarrow\rangle$, respectively. The final population errors in the target ground states are $4.4\times10^{-5}$, $1.24\times10^{-4}$, and $4.4\times10^{-5}$ in (a), (b), and (c), respectively, and their final phases are $0.0213$, $-1.5583$, and $0.0213$~rad, respectively.     }\label{Gate-2pulses-unitary}
\end{figure}

\begin{table*}[ht]
  \centering
  \begin{tabular}{|c|c|c|c|c|c|}
    \hline
      \multicolumn{2}{|c|}{}&First pulse&  \begin{tabular}{c}Second pulse \end{tabular} & \begin{tabular}{c}Third pulse \end{tabular}&Total duration \\   
   \hline  
  \multirow{2}{*} { Three-pulse gate}& Rabi frequency& $1.6088\Delta$ & $0.5932e^{i3\pi/4}\Delta$&$  1.6088 e^{i\beta/2}\Delta $& $3.14\pi/\Delta$  \\   
   \cline{2-5}  
   & pulse duration& $0.805\pi/\Delta$ & $1.532\pi/\Delta$&$ 0.805\pi/\Delta$& or $5.05\pi/\Omega_{\text{m}} $  \\   
   \hline 
   \multirow{2}{*} { Two-pulse gate}& Rabi frequency& $1.6088\Delta$ &  $  -1.6088  e^{i(\beta/2+\kappa)}\Delta $&\multirow{2}{*}{(not applicable)} & $1.61\pi/\Delta$  \\   
   \cline{2-4}  
   & pulse duration& $0.805\pi/\Delta$ & $ 0.805\pi/\Delta$& & or $2.59\pi/\Omega_{\text{m}} $  \\      
   \hline          
  \end{tabular}
  \caption{Parameters for creating nuclear-spin controlled-phase gate of the phase $\beta$. In the ideal blockade condition the gate map is diag$\{e^{-i\alpha} ,e^{-i\beta/2} , e^{-i\beta/2},e^{i\alpha}  \}$, where $\alpha$ is about 1.793~rad in the three-pulse gate, and is zero in the two-pulse gate. A CZ gate is realized via single-qubit rotations when choosing $\beta=\pi$ for both cases here. $\kappa$ is a phase to compensate an overall phase factor in the frame transform as discussed around Eq.~(\ref{kappaEquation}).   \label{table1}  }
\end{table*}

\section{Faster Gates assisted by Stark Shifts}\label{Sec03}
Here, we show that it is possible to realize the nuclear-spin gate in Sec.~\ref{Sec02} with only two global pulses when assisted by Stark shifts. There are two possible approaches to shifting the transition energy between the Rydberg state and the clock or ground state. First, one can shift the energy of the atomic states by usual methods. In Ref.~\cite{Barnes2022}, two nuclear spin qubit states with $m_I=-I, 1-I$ in $^{87}$Sr were isolated via Stark shifting other nuclear spin states away from the resonant beams. In Ref.~\cite{Jenkins2022}, light shifts in excited states were used to manipulate coherent control over the nuclear spin states in $^{171}$Yb. Second, the excitation of the inner electron of Rydberg states can be used. For example, energy shifts over $2\pi\times10$~MHz were demonstrated for both the $(6s6n)^3S_1$~\cite{Burgers2022} and the $(6s6n)^1S_0$~\cite{Pham2022} Rydberg states via the excitation of the Yb$^+6s\rightarrow6p_{1/2}$ transition of the inner electron in $^{174}$Yb. Extending the methods of Refs.~\cite{Burgers2022,Pham2022} to $^{171}$Yb would require calibrating polarization dependence of light shifts as in~\cite{Barnes2022,Jenkins2022}.   

To be consistent with Sec.~\ref{Sec02}, we suppose that during the first pulse of the gate we use a rotating frame that transfers the Hamiltonian $\hat{\mathbb{H}}$ to $e^{it\hat{R}}\hat{\mathbb{H}} e^{-it\hat{R}} - \hat{R}\equiv \hat{H}$ with
\begin{eqnarray}
 \hat{R} &=& \omega (\lvert r_+ \rangle\langle r_+|+ \lvert r_-  \rangle\langle r_-| ), \nonumber
\end{eqnarray}
where $\omega\pm\Delta=E_\pm$ is the energy separation~(divided by the reduced Planck constant) between the clock state and $\lvert r_\pm\rangle$, the energy is measured in reference to that of the clock state, and we ignore the energy separation between $\lvert \text{c}_\uparrow\rangle$ and $\lvert \text{c}_\downarrow\rangle$ since it is orders of magnitude smaller than the MHz-scale Rabi frequencies in a Gauss-scale magnetic field. In the second pulse, we assume that there are Stark shifts $\delta_\pm$ in $\lvert r_\pm\rangle$ with the condition $\delta_--\delta_+=4\Delta$, and another rotating frame 
\begin{eqnarray}
 \hat{R}' &=&( E_++\delta_++\Delta)\lvert r_+ \rangle\langle r_+|+ (E_-+\delta_--\Delta)\lvert r_-  \rangle\langle r_-|   \nonumber
\end{eqnarray}
is used. In this new frame, the state in Eq.~(\ref{Gate-p2}) becomes $
(e^{i\varphi_+'}\lvert r_+\text{c}_\downarrow\rangle + e^{i\varphi_-'}\lvert  \text{c}_\uparrow r_-\rangle)/\sqrt{2}$, where $\varphi_+'= \varphi_++\mathbb{T}_1(\delta_++2\Delta)$ and $\varphi_-'= \varphi_-+\mathbb{T}_1(\delta_--2\Delta)$. With the condition $\delta_--\delta_+=4\Delta$, one can see that $\varphi_+'- \varphi_+=\varphi_-'-\varphi_-$ which is $\kappa=\mathbb{T}_1(\delta_++2\Delta)$. In other words, the frame transfer changes the state in Eq.~(\ref{Gate-p2}) to
\begin{eqnarray}
e^{i\kappa}(e^{i\varphi_+}\lvert r_+\text{c}_\downarrow\rangle + e^{i\varphi_-}\lvert  \text{c}_\uparrow r_-\rangle)/\sqrt{2}, \label{kappaEquation}
\end{eqnarray}
while the four two-atom Rydberg states $\lvert r_\pm r_\pm\rangle$~(which appear during the pulse sequence when $V$ is finite) get an extra phase $2\kappa$ which is accounted for in the numerical simulation. 
By tuning the laser frequency to be at the middle between $\lvert r_-\rangle$ and $\lvert r_+\rangle$ and adding a phase $\pi+\beta/2+\kappa$ to the laser field, namely, 
\begin{eqnarray}
\Omega_2 = -e^{i(\beta/2+\kappa)}\Omega_1,\nonumber
\end{eqnarray}
the second pulse is with a Hamiltonian $[\Omega_{2}\lvert r_+\rangle\langle\text{c}_\uparrow\rvert  +\Omega_{2}\lvert r_-\rangle\langle\text{c}_\downarrow\rvert+\text{H.c.}]/2 -\Delta(\lvert r_+\rangle\langle r_+\rvert-\lvert r_-\rangle\langle r_-\rvert)$ after dipole and rotating wave approximations. After the second pulse which has the same duration as the first one, we realize Eq.~(\ref{Gate01}) with $\alpha=0$, and a CZ gate is realizable if $\beta=\pm\pi$ is chosen in the laser field. The total gate duration is $2\mathbb{T}_1\approx2.589\pi/\Omega_1$. A numerical simulation of the state dynamics for this two-pulse quantum gate is shown in Fig.~\ref{Gate-2pulses-unitary} by assuming $\delta_- =4\Delta$ and $\delta_+=0$~(a different choice for the Stark shift does not bring difference to the target gate map).

\begin{figure}
\includegraphics[width=2.5in]
{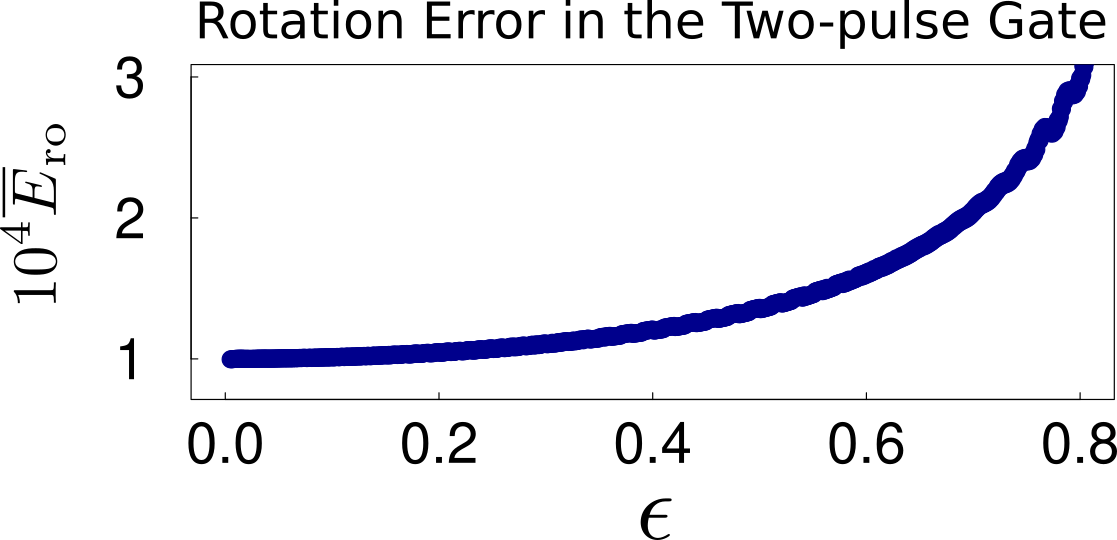}
 \caption{ Rotation error of the nuclear-spin gate~(rescaled by $10^4$) averaged by uniformly varying the Rydberg interaction $V$ in $[(1-\epsilon)V_0,~(1+\epsilon)V_0]$, where $V_0/2\pi=260$~MHz. Other parameters are the same as in Fig.~\ref{Gate-2pulses-unitary}. The gate fidelity $1- \overline{E}_{\text{ro}} -  E_{\text{decay}} $ is over $0.999$ for the $\epsilon$ shown here. }\label{Gate-rotation-2p}
\end{figure}

We turn to the analyses of the gate fidelity with finite Rydberg interactions. With $V=V_0$, numerical simulation shows that the gate map is diag$\{e^{i\nu} ,e^{i\mu} , e^{i\mu},e^{i\nu}  \}$, where $(\mu,\nu)\approx(-1.5583,0.02133)$~rad when $\beta = \pi$ and the state dynamics is shown in Fig.~\ref{Gate-2pulses-unitary}. The Rydberg superposition time is $t_{\text{Ryd}}\approx 1.47\pi/\Omega_1$ which leads to an error $E_{\text{decay}}=6.83\times10^{-4}$ with $\Omega=2\pi\times3.25$~MHz and $\tau=330~\mu$s. To characterize the error from the finiteness and fluctuation of $V$, we define 
\begin{eqnarray}
\hat{\mathbb{U}}= \text{diag} \{e^{i(\nu+\vartheta)} ,e^{i(\mu-\vartheta)},e^{i(\mu-\vartheta)},e^{i(\nu+\vartheta)} \}\label{GateV2p}
\end{eqnarray}
as the target gate, where $\vartheta =\pi/4- (\nu-\mu)/2\approx-0.004$~rad. By using the phase gates
\begin{eqnarray}
\lvert \text{c}_\uparrow\rangle&\rightarrowtail&  e^{-i(\nu+\vartheta)/2}\lvert \text{c}_\uparrow\rangle,\nonumber\\
\lvert \text{c}_\downarrow\rangle&\rightarrowtail& e^{i(\nu-2\mu+3\vartheta)/2} \lvert \text{c}_\downarrow\rangle,\nonumber
\end{eqnarray}
the map in Eq.~(\ref{GateV2p}) can be transformed to the CZ gate. We define the rotation error via Eq.~(\ref{GateF01}) with $\hat{U}$ replaced by the matrix in Eq.~(\ref{GateV2p}), and evaluate the average infidelity as in Eq.~(\ref{RoErrorV}), with the results shown in Fig.~\ref{Gate-rotation-2p} which shows that with a large relative fluctuation $\epsilon=0.8$ of $V$, the gate still has a large fidelity 0.999. One can see that compared to the three-pulse gate, the Rydberg-state-decay error is smaller here.

\section{Super Bell States}\label{Sec04}
The SBS in Eq.~(\ref{Bell}) can be prepared starting from the following two-atom state
\begin{eqnarray}
\lvert\psi(0)\rangle&=& |\text{cc}\rangle_{\text{e}}\otimes (\lvert\uparrow\uparrow\rangle_{\text{n}}+ \lvert\uparrow\downarrow\rangle_{\text{n}}+ \lvert\downarrow\uparrow\rangle_{\text{n}}+ \lvert\downarrow\downarrow\rangle_{\text{n}})/2,\label{initialState}
\end{eqnarray}
where the two-atom state is separable in that for each atom, the electronic state is in the optical clock state and the nuclear spin is in $(\lvert\uparrow\rangle_{\text{n}}+ \lvert\downarrow\rangle_{\text{n}})/\sqrt{2}$. Here, $\otimes$ is used because for good approximation, the electron and nuclear spin are decoupled in the ground~(g) and clock~(c) states ~\cite{Shi2021,Shi2021pra,Chen2022}. In Eq.~(\ref{Bell}), $(\theta,\theta')$ are dependent on the Rydberg interactions and they approach $(0,\pi)$ in the ideal blockade condition. Because $_{\text{e}}\langle \text{cc}\rvert\Phi\rangle_{\text{e}}=~_{\text{n}}\langle \Psi\rvert\Phi\rangle_{\text{n}} =0   $, one can define a pair of orthogonal two-atom electronic states 
\begin{eqnarray}
  \lvert +\rangle_{\text{e}}  &\equiv& \lvert \text{cc}\rangle_{\text{e}} , \nonumber\\
 \lvert-\rangle_{\text{e}}  &\equiv& \lvert\Phi\rangle_{\text{n}} ,  \nonumber
\end{eqnarray} 
and a pair of orthogonal two-atom nuclear-spin states 
\begin{eqnarray} 
  \lvert +\rangle_{\text{n}}  &\equiv& \lvert\Phi\rangle_{\text{n}} , \nonumber\\
 \lvert-\rangle_{\text{n}}  &\equiv& \lvert\Psi\rangle_{\text{n}} ,\nonumber
\end{eqnarray}
so that Eq.~(\ref{Bell}) can also be written as
\begin{eqnarray}
 |\text{SBS}\rangle
 &\equiv&\frac{1}{\sqrt{2}}\bigg(\lvert+\rangle_{\text{e}} \otimes \lvert+\rangle_{\text{n}}   + \lvert-\rangle_{\text{e}}  \otimes \lvert-\rangle_{\text{n}} \bigg),\label{Bell4}
\end{eqnarray}
which is a Bell state entangled between the electronic and nuclear spin states in two atoms. Combining Eqs.~(\ref{Bell2}),~(\ref{Bell3}), and~(\ref{Bell4}), one can find that SBS is a ``large'' Bell state formed with three ``smaller'' Bell states and a product state $\lvert \text{cc}\rangle_{\text{e}}$.

\subsection{A three-pulse protocol}\label{Sec03A}
We describe a three-pulse sequence for creating SBS. 

\subsubsection{The first pulse}\label{sec-pulse1}
First, with an ultra-violet~(UV) laser excitation of pulse duration $\mathbb{T}_{\text{p}1}^{(\text{\tiny{S}})}$, 
\begin{eqnarray}
 &&\lvert \text{c}_\uparrow \rangle \xrightarrow[\text{detuned by~} \Delta]{\Omega_{\text{c}\uparrow}=\Omega^{(\text{S})} } \lvert r_+  \rangle ,\nonumber\\
 &&\lvert \text{c}_\downarrow \rangle \xrightarrow[\text{detuned by~} -\Delta]{\Omega_{\text{c}\downarrow}= \Omega^{(\text{S})}}  \lvert r_-  \rangle , \nonumber
\end{eqnarray}
the following state transfer is realized,
\begin{eqnarray}
 &&\lvert \text{c}_\uparrow \text{c}_\uparrow\rangle \xrightarrow{ } e^{-i\hat{H}\mathbb{T}_{\text{p}1}^{(\text{\tiny{S}})} }\lvert \text{c}_\uparrow \text{c}_\uparrow\rangle,\nonumber\\
&& \lvert \text{c}_\downarrow \text{c}_\downarrow\rangle \xrightarrow{ } e^{-i\hat{H}\mathbb{T}_{\text{p}1}^{(\text{\tiny{S}})} } \lvert \text{c}_\downarrow \text{c}_\downarrow\rangle ,\nonumber\\
&& \lvert \text{c}_\uparrow \text{c}_\downarrow\rangle \xrightarrow{ }  \lvert \text{c}_\uparrow \text{c}_\downarrow\rangle ,\nonumber\\
&& \lvert \text{c}_\downarrow \text{c}_\uparrow\rangle \xrightarrow{ }  \lvert \text{c}_\downarrow \text{c}_\uparrow\rangle, \label{pulse1}
\end{eqnarray}
where $\hat{H}$ is the Hamiltonian shown in Appendix~\ref{AppendixA}. In the blockade regime, one can easily find, according to the picture of generalized Rabi oscillation~\cite{Shi2017,Shi2018prapp,Shi2019prap,Shi2020}, that the last two transitions in Eq.~(\ref{pulse1}) can be realized with a pulse duration $\mathbb{T}_{\text{p}1}^{(\text{\tiny{S}})}=2\pi/\sqrt{\Delta^2+0.5\Omega^2}$, where the ratio between $\Delta$ and $\Omega$ is determined later in Sec.~\ref{sec-pulse2} below.

\subsubsection{The second pulse}\label{sec-pulse2} 
Second, with a two-photon Rydberg excitation of pulse duration $\mathbb{T}_{\text{p}2}^{(\text{\tiny{S}})}=2\pi/\sqrt{\Delta^2+0.5\eta^2\Omega^2}$,  
\begin{eqnarray}
 &&\lvert \text{c}_\uparrow \rangle \xrightarrow[\text{detuned by~} \Delta]{\Omega_{\text{c}\uparrow}=\eta\Omega^{(\text{S})}}  \lvert r_+  \rangle ,\nonumber\\
 &&\lvert \text{c}_\downarrow \rangle \xrightarrow[\text{detuned by~} -\Delta]{\Omega_{\text{c}\downarrow}= \eta\Omega^{(\text{S})}}  \lvert r_-  \rangle , \nonumber
\end{eqnarray}
the following state transfer is realized,
\begin{eqnarray}
 &&e^{-i\hat{H}\mathbb{T}_{\text{p}1}^{(\text{\tiny{S}})} }\lvert \text{c}_\uparrow \text{c}_\uparrow\rangle \xrightarrow{ } e^{i\alpha}\frac{\lvert r_+ \text{c}_\uparrow\rangle+\lvert \text{c}_\uparrow r_+\rangle}{\sqrt{2}},\nonumber\\
&& e^{-i\hat{H}\mathbb{T}_{\text{p}1}^{(\text{\tiny{S}})} }\lvert \text{c}_\downarrow \text{c}_\downarrow\rangle \xrightarrow{ }e^{i(\pi-\alpha)}\frac{\lvert r_- \text{c}_\downarrow\rangle +\lvert \text{c}_\downarrow r_-\rangle}{\sqrt{2}} ,\nonumber\\
&& \lvert \text{c}_\uparrow \text{c}_\downarrow\rangle \xrightarrow{ }  \lvert \text{c}_\uparrow \text{c}_\downarrow\rangle ,\nonumber\\
&& \lvert \text{c}_\downarrow \text{c}_\uparrow\rangle \xrightarrow{ }  \lvert \text{c}_\downarrow \text{c}_\uparrow\rangle, \label{pulse2}
\end{eqnarray}
where the Hamiltonian is still in the form as in Appendix~\ref{AppendixA}. We find that with 
\begin{eqnarray}
 |\Omega^{(\text{S})}|/ \Delta&=& 1.608, \nonumber\\
 \eta&=& -0.4606, \label{condition-p1p2}
\end{eqnarray}
the transitions in Eqs.~(\ref{pulse1}) and~(\ref{pulse2}) can be realized perfectly when $V/\Delta$ is infinite, where $\alpha=-0.8502$~(2.291)~rad when $\Omega^{(\text{S})}$ is positive~(negative). Errors due to finite blockade will be analyzed later.

\subsubsection{The third pulse}\label{sec-pulse3}
Before going to the third pulse, it is useful to point out that the description from Sec.~\ref{sec-pulse1} to Sec.~\ref{sec-pulse2} is in one rotating frame, and a different rotating frame is used in the third pulse. Details for this change of rotating frames is given in Appendix~\ref{AppendixB}, which shows that the frame transform changes the phase $\alpha$ in the first two equations of Eq.~(\ref{pulse2}) to
\begin{eqnarray}
\alpha'&=&\alpha+\Delta\left[\mathbb{T}_{\text{p}1}^{(\text{\tiny{S}})}+\mathbb{T}_{\text{p}2}^{(\text{\tiny{S}})}\right]. \nonumber
\end{eqnarray}

The third pulse is realized by two-photon Rydberg laser excitation of the transitions
\begin{eqnarray}
 &&\lvert \text{g}_\uparrow \rangle \xrightarrow[]{\Omega_{\text{gr}\uparrow} = \Omega_{\text{eff}}(1+e^{2it\Delta})}  \lvert r_+  \rangle ,\nonumber\\
 &&\lvert \text{g}_\downarrow \rangle \xrightarrow[]{\Omega_{\text{cr}\downarrow} = -\Omega_{\text{eff}}(1+e^{-2it\Delta})}  \lvert r_-  \rangle , \nonumber\label{ground-Rydberg01}
\end{eqnarray}
i.e., two sets of laser fields of equal strength are sent, one resonant with $\lvert \text{g}_\uparrow \rangle\leftrightarrow  \lvert r_+  \rangle$, the other resonant with $\lvert \text{g}_\downarrow \rangle\leftrightarrow  \lvert r_-  \rangle$, each with a Rabi frequency $\Omega_{\text{eff}}$ where eff denotes that it is derived by adiabatic elimination of the intermediate states~\cite{Shi2021}. According to Eq.~(\ref{pulse2}), there is no Rydberg population in the states evolved from the input states $\lvert \text{c}_\uparrow \text{c}_\downarrow\rangle$ and $\lvert \text{c}_\downarrow \text{c}_\uparrow\rangle$, and only the states evolved from the two input states $\lvert \text{c}_\uparrow \text{c}_\uparrow\rangle$ and $\lvert \text{c}_\downarrow \text{c}_\downarrow\rangle$ respond to the third pulse. For the input state $\lvert \text{c}_\uparrow \text{c}_\uparrow\rangle$ which is  $e^{i\alpha'}\frac{\lvert r_+ \text{c}_\uparrow\rangle+\lvert \text{c}_\uparrow r_+\rangle}{\sqrt{2}}$ at the beginning of the third pulse, the Hamiltonian is
\begin{eqnarray}
 \hat{H}_{\uparrow\uparrow} &=&  \frac{\Omega_{\text{g}\uparrow}}{2} \frac{\lvert r_+ \text{c}_\uparrow\rangle+\lvert \text{c}_\uparrow r_+\rangle}{\sqrt{2}}   \frac{\langle \text{g}_\uparrow\text{c}_\uparrow\rvert+\langle \text{c}_\uparrow \text{g}_\uparrow\rvert}{\sqrt{2}} +\text{H.c.}\nonumber
\end{eqnarray}
For the input state $\lvert \text{c}_\downarrow \text{c}_\downarrow\rangle$ which is  $e^{i(\pi-\alpha')}\frac{\lvert r_- \text{c}_\downarrow\rangle +\lvert \text{c}_\downarrow r_-\rangle}{\sqrt{2}}$, we have
\begin{eqnarray}
 \hat{H}_{\downarrow\downarrow} &=&  \frac{\Omega_{\text{g}\downarrow}}{2} \frac{\lvert r_- \text{c}_\downarrow\rangle +\lvert \text{c}_\downarrow r_-\rangle}{\sqrt{2}}   \frac{\langle \text{g}_\downarrow\text{c}_\downarrow\rvert+\langle \text{c}_\downarrow \text{g}_\downarrow\rvert}{\sqrt{2}} +\text{H.c.}\nonumber
\end{eqnarray}
We have also shown the above Hamiltonians without using the collective entangled basis states in Appendix~\ref{AppendixC}. We find that with 
\begin{eqnarray}
 |\Omega_{\text{eff}}|/ \Delta&=& 0.7064, \label{condition-p3}
\end{eqnarray}
the third pulse leads to the state transfer 
\begin{eqnarray}
 &&  e^{i\alpha'}\frac{\lvert r_+ \text{c}_\uparrow\rangle+\lvert \text{c}_\uparrow r_+\rangle}{\sqrt{2}}\xrightarrow{ } e^{i(\alpha'+\beta)}\frac{\lvert \text{g}_\uparrow  \text{c}_\uparrow\rangle+\lvert \text{c}_\uparrow \text{g}_\uparrow\rangle }{\sqrt{2}}  ,\nonumber\\
&&   e^{i(\pi-\alpha')}\frac{\lvert r_- \text{c}_\downarrow\rangle +\lvert \text{c}_\downarrow r_-\rangle}{\sqrt{2}}\xrightarrow{ }  e^{i(\pi-\alpha'-\beta)}\frac{\lvert \text{g}_\downarrow   \text{c}_\downarrow\rangle +\lvert \text{c}_\downarrow  \text{g}_\downarrow  \rangle}{\sqrt{2}}, \nonumber\\\label{pulse3}
\end{eqnarray}
with a pulse duration 
\begin{eqnarray}
  \mathbb{T}_{\text{p}3}^{(\text{\tiny{S}})}= \frac{1.688\pi}{\Delta},\nonumber
\end{eqnarray}
while the states $\lvert \text{c}_\uparrow \text{c}_\downarrow\rangle$ and $\lvert \text{c}_\downarrow \text{c}_\uparrow\rangle$ don't evolve. We have $\beta= -1.575$~rad when $\Omega_{\text{eff}}$ is positive, and $\beta= 1.567$ when $\Omega_{\text{eff}}$ is negative.   

\begin{figure}
\includegraphics[width=3.2in]
{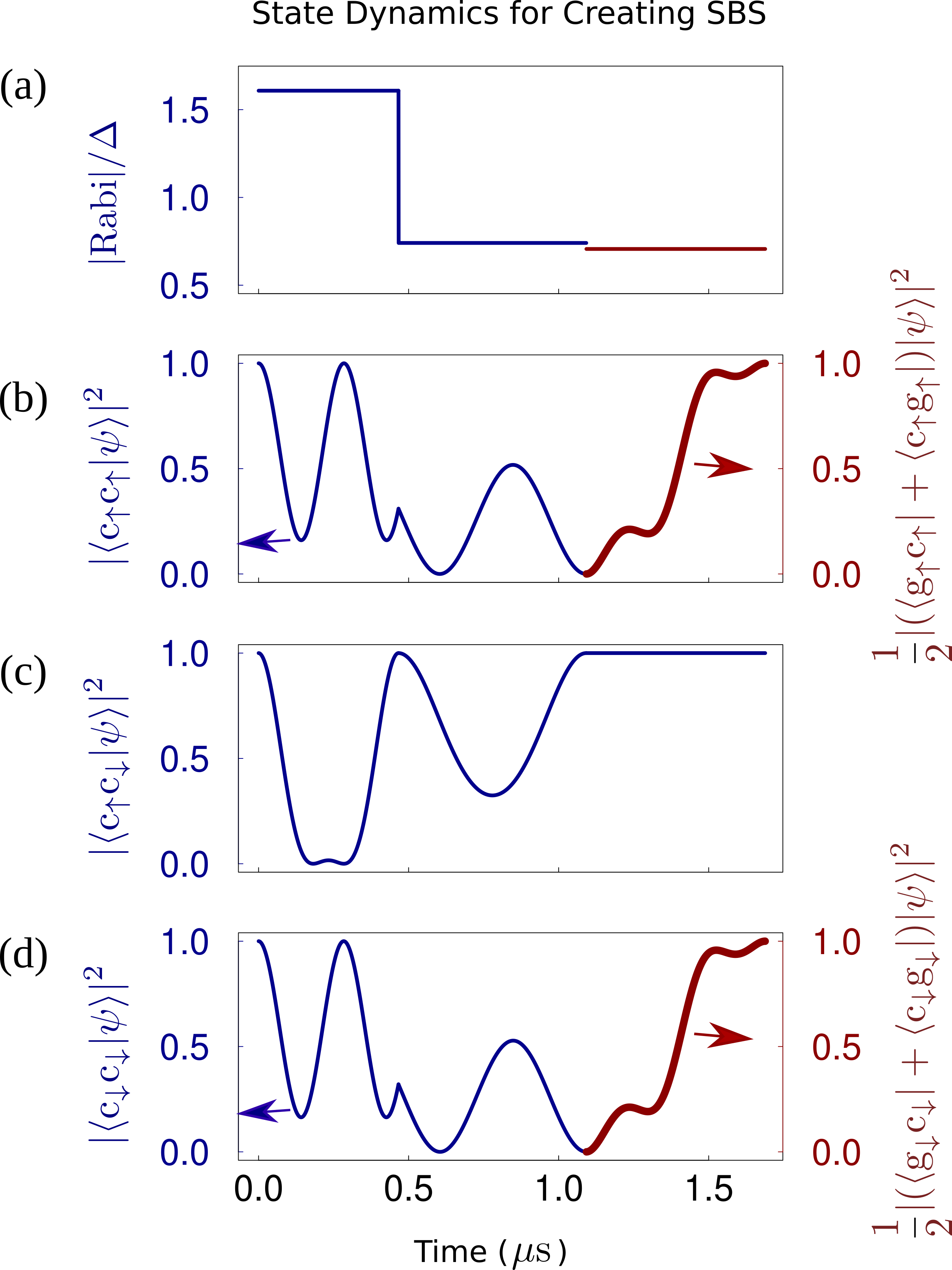}
 \caption{State dynamics of the SBS protocol for different input states with $V_0/2\pi=260$~MHz and maximal Rabi frequency $\Omega^{(\text{S})}/2\pi\approx2.28$~MHz when ignoring Rydberg-state decay. (a) Magnitudes of Rabi frequencies in units of $\Delta$. Here, the third pulse is with two-photon laser excitation, where one laser is resonant for the transition $\lvert \text{g}_\uparrow \rangle\rightarrow \lvert r_+  \rangle$ and the other laser resonant for $ \lvert \text{g}_\downarrow \rangle \rightarrow\lvert r_-  \rangle$, each with a Rabi frequency $\Omega_{\text{eff}}=2\pi\times1$~MHz. These two lasers can be from one laser source via pulse pickers for shifting the frequency of one sub-beam by $2\Delta$. (b), (c), and (d) show the population of the ground-state component in the wavefunction when the input states for the SBS protocol are $\lvert \text{c}_\uparrow \text{c}_\uparrow\rangle$, $\lvert \text{c}_\uparrow \text{c}_\downarrow\rangle$, and $\lvert \text{c}_\downarrow \text{c}_\downarrow\rangle$, respectively. The thin~(thick) curves in (b) and (d) denote population in the product~(Bell) states. For the target ground states $\lvert\Phi\rangle_{\text{e}}  \otimes \lvert\uparrow\uparrow\rangle_{\text{n}}$ in (b), $\lvert\text{c}_\uparrow\text{c}_\downarrow\rangle$ in (c), and $\lvert\Phi\rangle_{\text{e}}  \otimes \lvert\downarrow\downarrow\rangle_{\text{n}}$ in (d), the final populations are $1-1.4\times10^{-4}$, $1-3.1\times10^{-5}$, and $1-5.5\times10^{-5}$ respectively, and the final phases are $\Delta\left[\mathbb{T}_{\text{p}1}^{(\text{\tiny{S}})}+\mathbb{T}_{\text{p}2}^{(\text{\tiny{S}})}\right]-2.406$~rad, $0.009$~rad, and $-\Delta\left[\mathbb{T}_{\text{p}1}^{(\text{\tiny{S}})}+\mathbb{T}_{\text{p}2}^{(\text{\tiny{S}})}\right]-0.697$~rad, respectively. These are different from those described in Secs.~\ref{sec-pulse2} and~\ref{sec-pulse3} for the Rydberg interaction is finite~(its fluctuation is studied in Sec.~\ref{sec03B}).    }\label{SBS-unitary}
\end{figure}

By defining nuclear spin up and down states with 
\begin{eqnarray}
 \text{new}\lvert \text{g}_\uparrow  \rangle&=&   e^{i(\alpha'+\beta)/2} \lvert \text{g}_\uparrow   \rangle  ,\nonumber\\
 \text{new}\lvert \text{c}_\uparrow  \rangle&=&   e^{i(\alpha'+\beta)/2} \lvert \text{c}_\uparrow   \rangle    ,\nonumber\\
 \text{new}\lvert \text{g}_\downarrow  \rangle &=&   e^{-i(\alpha'+\beta)/2} \lvert \text{g}_\downarrow   \rangle   ,\nonumber\\
 \text{new}\lvert \text{c}_\downarrow \rangle  &=&   e^{-i(\alpha'+\beta)/2} \lvert \text{c}_\downarrow   \rangle   , \label{basisDef-SBS}
\end{eqnarray}
the two final states in Eq.~(\ref{pulse3}) are written as
\begin{eqnarray}
 &&   \frac{\lvert \text{g}_\uparrow  \text{c}_\uparrow\rangle+\lvert \text{c}_\uparrow \text{g}_\uparrow\rangle }{\sqrt{2}}  ,\nonumber\\
&&   -\frac{\lvert \text{g}_\downarrow   \text{c}_\downarrow\rangle +\lvert \text{c}_\downarrow  \text{g}_\downarrow  \rangle}{\sqrt{2}}. \nonumber
\end{eqnarray}
Looking through Eqs.~(\ref{pulse1}),~(\ref{pulse2}),~(\ref{pulse3}), and the basis definition in Eq.~(\ref{basisDef-SBS}), one can see that 
\begin{eqnarray}
 &&\lvert \text{c}_\uparrow \text{c}_\uparrow\rangle \rightarrowtail{ }  \frac{\lvert \text{g}_\uparrow  \text{c}_\uparrow\rangle+\lvert \text{c}_\uparrow \text{g}_\uparrow\rangle }{\sqrt{2}} =   \frac{\lvert \text{cg}\rangle+\lvert  \text{gc}\rangle }{\sqrt{2}} \otimes\lvert \uparrow\uparrow\rangle ,\nonumber\\
&& \lvert \text{c}_\downarrow \text{c}_\downarrow\rangle \rightarrowtail  -\frac{\lvert \text{g}_\downarrow   \text{c}_\downarrow\rangle +\lvert \text{c}_\downarrow  \text{g}_\downarrow  \rangle}{\sqrt{2}} =  - \frac{\lvert \text{cg}\rangle+\lvert  \text{gc}\rangle }{\sqrt{2}} \otimes\lvert \downarrow\downarrow\rangle,\nonumber\\
&& \lvert \text{c}_\uparrow \text{c}_\downarrow\rangle \rightarrowtail  \lvert \text{c}_\uparrow \text{c}_\downarrow\rangle= \lvert \text{cc}\rangle_{\text{e}} \otimes\lvert \uparrow\downarrow\rangle ,\nonumber\\
&& \lvert \text{c}_\downarrow \text{c}_\uparrow\rangle \rightarrowtail \lvert \text{c}_\downarrow \text{c}_\uparrow\rangle= \lvert  \text{cc}\rangle_{\text{e}} \otimes\lvert \downarrow\uparrow\rangle , \label{IdealGate}  
\end{eqnarray}
is realized with a total duration $
 \sum_{x=1}^3\mathbb{T}_{\text{p}x}^{(\text{\tiny{S}})} $ which is about $4.781\pi/\Delta$, or $7.687\pi/|\Omega^{(\text{S})}|$, or $3.377\pi/|\Omega_{\text{eff}}|$ with the relations shown in Eqs.~(\ref{condition-p1p2}) and~(\ref{condition-p3}). If we start from the state in Eq.~(\ref{initialState}), the transform in Eq.~(\ref{IdealGate}) leads to SBS in Eq.~(\ref{Bell}). A numerical simulation with a practical $V$ is shown in Fig.~\ref{SBS-unitary}.

\subsubsection{Another SBS}
Using a similar strategy as in Secs.~\ref{sec-pulse1}-\ref{sec-pulse3}, one can find that starting from 
\begin{eqnarray}
\lvert\psi(0)\rangle&=& |\text{gg}\rangle_{\text{e}}\otimes (\lvert\uparrow\uparrow\rangle_{\text{n}}+ \lvert\uparrow\downarrow\rangle_{\text{n}}+ \lvert\downarrow\uparrow\rangle_{\text{n}}+ \lvert\downarrow\downarrow\rangle_{\text{n}})/2,\nonumber\label{initialState-2}
\end{eqnarray}
the following SBS 
\begin{eqnarray}
 |\text{SBS}\rangle'
 &\equiv&\frac{1}{\sqrt{2}}\bigg(\lvert\text{gg}\rangle_{\text{e}} \otimes \lvert\Phi\rangle_{\text{n}}   + \lvert\Phi\rangle_{\text{e}}  \otimes \lvert\Psi\rangle_{\text{n}} \bigg)  \label{Bell-2}
\end{eqnarray}
can be prepared by following the three pulses in Secs.~\ref{sec-pulse1},~\ref{sec-pulse2}, and~\ref{sec-pulse3} with the UV laser excitation and the two-photon ground-Rydberg excitation exchanged.  

The state in Eq.~(\ref{Bell-2}) can also be prepared from the SBS in Eq.~(\ref{Bell})~(with a trivial overall $\pi$ phase). The method is simple, by using the ground-clock state transition
\begin{eqnarray}
  \lvert\text{g}\rangle_{\text{e}} \leftrightarrow \lvert\text{c}\rangle_{\text{e}}\label{g-c-transition}
\end{eqnarray}
with a duration $\pi/\Omega_{\text{gc}}$, where $\Omega_{\text{gc}}$ is the Rabi frequency for the ground-clock transition, $\lvert\text{cc}\rangle_{\text{e}}\rightarrow- \lvert\text{gg}\rangle_{\text{e}}$ and $\lvert\Phi\rangle_{\text{e}}\rightarrow-\lvert\Phi\rangle_{\text{e}}$ are realized~\cite{Shi2018prapp}, leading to $|\text{SBS}\rangle\rightarrow-|\text{SBS}\rangle'$. As analyzed in Ref.~\cite{Chen2022}, $\Omega_{\text{gc}}=2\pi\times0.2$~MHz can be realized, with which Eq.~(\ref{Bell-2}) can be formed from Eq.~(\ref{Bell}) with a pulse duration $2.5~\mu$s.

\subsection{Numerical analyses}\label{sec03B}
Here, we numerically study the fidelity for creating SBS with the pulses of Secs.~\ref{sec-pulse1},~\ref{sec-pulse2}, and~\ref{sec-pulse3}. Three main factors limit the fidelity, the Rydberg state decay, the finiteness of $V$, and the fluctuation of $V$.  

The Rydberg state decay is inversely proportional to the Rabi frequencies in the protocol. The ground-Rydberg transition requires two-photon excitation, for which the Rabi frequency can be small. As analyzed in Ref.~\cite{Shi2021}, $\Omega_{\text{eff}}/2\pi=1.4$~MHz can be realized for exciting a Rydberg state with principal quantum number $n\sim 70$. During the third pulse of Sec.~\ref{sec-pulse3}, two sets of two-photon transitions shall be used, one resonant with $\lvert r_+\rangle$, and the other resonant with $\lvert r_-\rangle$. To realize such transitions with one laser source for the ground-intermediate state transition, and another laser source for the intermediate-Rydberg state transition, the transition from the intermediate state to $\lvert r_+\rangle$ and $\lvert r_-\rangle$ can be realized by dividing the laser beam to two halves for which the frequency of one half shall be shifted by $2\Delta$ via pulse pickers. This is experimentally feasible as shown in Ref.~\cite{Barnes2022}, where one laser source was used to drive a two-photon Raman transition assisted by several electro-optic modulators and crossed acousto-optic deflectors. This means that if we employ two lasers of similar powers in the analyses of Ref.~\cite{Shi2021}, the available $\Omega_{\text{eff}}$ will be reduced by a factor of $\sqrt{2}$. For this reason, we assume $\Omega_{\text{eff}}/2\pi=1$~MHz, which corresponds to $2\Delta/2\pi=2.83$~MHz~(realized with a B-field of 1.5~G) according to Eq.~(\ref{condition-p3}). By Eq~(\ref{condition-p1p2}), the UV laser Rabi frequencies are $\Omega^{(\text{S})}/2\pi\approx2.28$~MHz for the first pulse, and $|\eta|\Omega^{(\text{S})}/2\pi\approx1.05$~MHz for the second pulse. Concerning that the UV laser Rabi frequency can be quite large~\cite{Chen2022}, the above analyses show that the speed for creating SBS is limited by the smaller two-photon ground-Rydberg Rabi frequency. Note, however, that the analyses in Ref.~\cite{Shi2021} were quite conservative, and in principle higher ground-Rydberg Rabi frequencies should be achievable. The error due to Rydberg-state decay is given by Eq.~(\ref{decayE}),
where $t_{\text{Ryd}}$ is the Rydberg superposition time. We numerically find $ t_{\text{Ryd}} = 0.55\mathbb{T}_{\text{p}1} +0.57 \mathbb{T}_{\text{p}2} +0.23 \mathbb{T}_{\text{p}3}$. 
With $\Omega_{\text{eff}}/2\pi=1$~MHz, and $\Omega^{(\text{S})}$ and $\Delta$ given by Eqs.~(\ref{condition-p1p2}) and~(\ref{condition-p3}), we find $t_{\text{Ryd}} =0.75~\mu$s, leading to $t_{\text{Ryd}}=2.2\times10^{-3}$ with $\tau=330~\mu$s~\cite{Shi2021}.

\begin{figure}
\includegraphics[width=2.5in]
{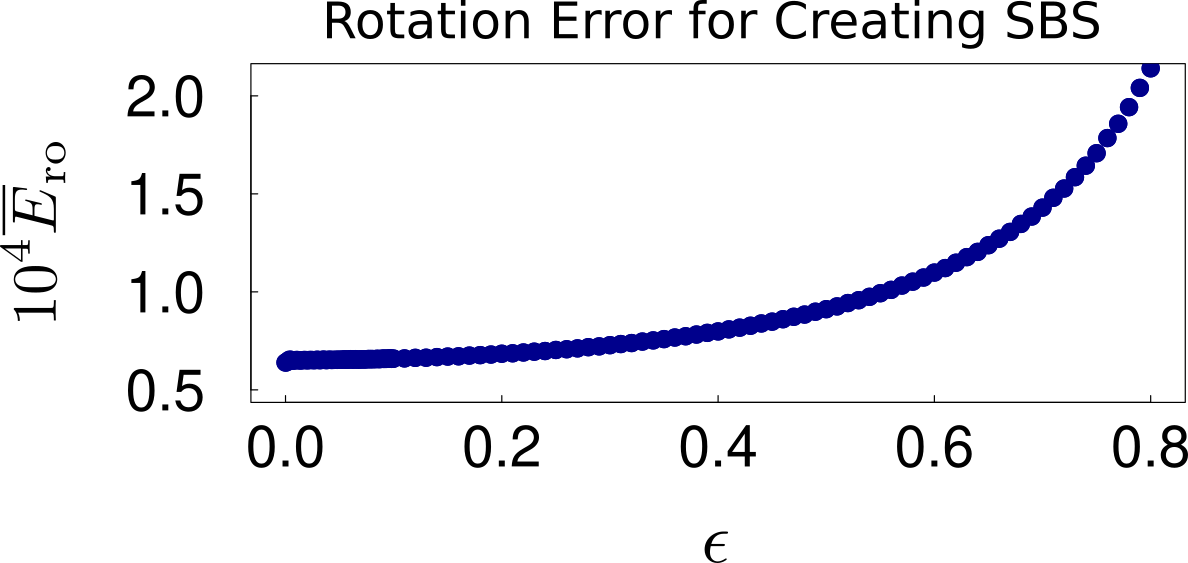}
 \caption{ Rotation error~(rescaled by $10^4$) of Eq.~(\ref{rotation-def}) averaged by uniformly varying the Rydberg interaction $V$ in $[(1-\epsilon)V_0,~(1+\epsilon)V_0]$ for creating SBS. Here, $V_0/2\pi=260$~MHz, $\Omega_{\text{eff}}/2\pi=1$~MHz, and $\Omega^{(\text{S})}$ and $\Delta$ are given by Eqs.~(\ref{condition-p1p2}) and~(\ref{condition-p3}). The fidelity of SBS generation, $1- \overline{E}_{\text{ro}} -  E_{\text{decay}} $, is about $0.998$ dominated by the Rydberg-state decay. }\label{SBS-rotation}
\end{figure}
The analyses in Sec.~\ref{sec03B} assumed infinite Rydberg interactions, but a finite $V$ leads to population loss to the computational basis states. Moreover, it leads to some population of the two-atom Rydberg states $\lvert r_+ r_+\rangle,\lvert r_+ r_-\rangle,\lvert r_- r_+\rangle$, and $\lvert r_- r_-\rangle$ during the laser excitation, which alters the dynamics a little bit and finally leads to phases in the computational states different compared to those in Eq.~(\ref{pulse3}). In other words, the final SBS with a certain $V$ is no longer
\begin{eqnarray}
 &&\frac{1}{2}\Big[\frac{\lvert \text{cg}\rangle+\lvert  \text{gc}\rangle }{\sqrt{2}} \otimes\left(e^{i(\alpha'+\beta)}\lvert \uparrow\uparrow\rangle_{\text{n}}+ e^{i(\pi-\alpha'-\beta )} \lvert \downarrow\downarrow\rangle_{\text{n}}\right)\nonumber\\
&& +\lvert \text{cc}\rangle_{\text{e}} \otimes\left(\lvert \uparrow\downarrow\rangle_{\text{n}} +\lvert \downarrow\uparrow\rangle_{\text{n}}\right) \Big] , \nonumber
\end{eqnarray}
but
\begin{eqnarray}
 &&\frac{1}{2}\Big[\frac{\lvert \text{cg}\rangle+\lvert  \text{gc}\rangle }{\sqrt{2}} \otimes\left(\chi_1 e^{i\theta_1}\lvert \uparrow\uparrow\rangle_{\text{n}}+\chi_2 e^{i\theta_2} \lvert \downarrow\downarrow\rangle_{\text{n}}\right)\nonumber\\
&& +\chi_3 e^{i\theta_3}\lvert \text{cc}\rangle_{\text{e}} \otimes\left(\lvert \uparrow\downarrow\rangle_{\text{n}} +\lvert \downarrow\uparrow\rangle_{\text{n}}\right) \Big], \nonumber
\end{eqnarray}
where $\theta_1,~\theta_2$, and $\theta_3$ are dependent on $V$, and $\chi_{1},\chi_{2},\chi_{3}$ should be 1 in the ideal case, but they are smaller than 1 due to the population loss when $V$ is finite. To investigate the rotation error with finite $V$, we define
\begin{eqnarray}
 \lvert \mathbb{SBS} \rangle&=&\frac{1}{2}\Big[\frac{\lvert \text{cg}\rangle+\lvert  \text{gc}\rangle }{\sqrt{2}} \otimes\left(e^{i\theta_1}\lvert \uparrow\uparrow\rangle_{\text{n}}+ e^{i\theta_2} \lvert \downarrow\downarrow\rangle_{\text{n}}\right)\nonumber\\
&& + e^{i\theta_3}\lvert \text{cc}\rangle_{\text{e}} \otimes\left(\lvert \uparrow\downarrow\rangle_{\text{n}} +\lvert \downarrow\uparrow\rangle_{\text{n}}\right) \Big], \label{SBS-final}
\end{eqnarray}
as the target SBS. Since a global phase is insignificant, Eq.~(\ref{SBS-final}) can be written as Eq.~(\ref{Bell}) where $\theta=\theta_1-\theta_3$ and $\theta'=\theta_2-\theta_3$ in Eq.~(\ref{Bell2}). The reason for the phase $\theta_1$ of $\frac{\lvert \text{cg}\rangle+\lvert  \text{gc}\rangle }{\sqrt{2}} \otimes\lvert \uparrow\uparrow\rangle $ to differ from $\theta_2$ of $\frac{\lvert \text{cg}\rangle+\lvert  \text{gc}\rangle }{\sqrt{2}} \otimes\lvert \downarrow\downarrow\rangle $ is that the detunings $V\pm2\Delta$ for them are different, shown in Eqs.~(\ref{Huu}) and~(\ref{Hdd}), which result in different state dynamics. With $V_0=2\pi\times260$~MHz, we find
\begin{eqnarray}
 \theta_1&=&-2.406+\Delta(\mathbb{T}_{\text{p}1}+\mathbb{T}_{\text{p}2}),\nonumber\\
 \theta_2&=&-0.697-\Delta(\mathbb{T}_{\text{p}1}+\mathbb{T}_{\text{p}2}),\nonumber\\
 \theta_3&=&0.009.\nonumber
\end{eqnarray}
By defining new nuclear-spin and clock states~(or, using single-qubit phase gates),
\begin{eqnarray}
 \text{new}\lvert \uparrow\rangle_{\text{n}} & =& e^{i(2\theta_3-3\theta_1-\theta_2)/4}  \lvert \uparrow\rangle_{\text{n}},\nonumber\\
\text{new} \lvert \downarrow\rangle_{\text{n}} & =&  e^{i(2\theta_3-\theta_1-3\theta_2)/4}  \lvert \downarrow\rangle_{\text{n}},\nonumber\\
 \text{new} \lvert \text{c}\rangle_{\text{e}} & =&   e^{i(\theta_1+\theta_2-2\theta_3)/2}  \lvert \text{c}\rangle_{\text{e}},\nonumber
\end{eqnarray}
Eq.~(\ref{SBS-final}) can be written as
\begin{eqnarray}
 &&\frac{\lvert \text{cg}\rangle_{\text{e}}+\lvert  \text{gc}\rangle_{\text{e}} }{2\sqrt{2}} \otimes\left( \lvert \uparrow\uparrow\rangle_{\text{n}}  +  \lvert \downarrow\downarrow\rangle_{\text{n}} \right)+\frac{\lvert \text{cc}\rangle_{\text{e}}  }{2}\otimes\left(\lvert \uparrow\downarrow\rangle_{\text{n}}  +\lvert \downarrow\uparrow\rangle_{\text{n}} \right) . \nonumber
\end{eqnarray}
However, the target state is still Eq.~(\ref{SBS-final}), in which the angles $\theta_k$, $k=1,2,3$ fluctuate when $V$ fluctuates due to the spatial fluctuation of the two atoms, leading to a rotation error,
\begin{eqnarray}
 E_{\text{ro}} &=& 1-\langle \mathbb{SBS} |\hat{\rho}|\mathbb{SBS} \rangle , \label{rotation-def}
\end{eqnarray}
where $\hat{\rho}$ is the density matrix of the actual state realized with the three-pulse protocol. Because the Rydberg-state decay is analytically approximated above so as to get an upper bound for the error~\cite{Saffman2005}, we use unitary dynamics to simulate the blockade error so that Eq.~(\ref{rotation-def}) reduces to $1-\lvert\langle  \psi|\mathbb{SBS} \rangle |^2$ where $\lvert \psi\rangle$ is the final state evolved from Eq.~(\ref{initialState}). To evaluate the rotation error, we consider the average $ \overline{E}_{\text{ro}} = \int E_{\text{ro}}(V) dV/\int dV$, where the integration is over $V\in[(1-\epsilon)V_0,~(1+\epsilon)V_0]$. To have a lower bound for the fidelity, we consider a uniform distribution of $V$ instead of a Gaussian distribution. The numerical results for different $\epsilon$ are shown in Fig.~\ref{SBS-rotation}. We find that for $\epsilon<0.55$, the average rotation error $\overline{E}_{\text{ro}} $ is smaller than $10^{-4}$. Remarkably, even with $\epsilon=0.8$, $\overline{E}_{\text{ro}} $ is $2.14\times10^{-4}$. This robustness benefits from the intrinsic property of the blockade mechanism~\cite{Shi2021qst}. The fidelity $\mathcal{F} = 1- \overline{E}_{\text{ro}} -  E_{\text{decay}}$ for creating SBS is about 99.8$\%$ for $\epsilon$ up to $0.8$, which is dominated by the Rydberg-state decay. This fidelity should be possible to be improved if larger ground-Rydberg Rabi frequency is available.  
\begin{table*}[ht]
  \centering
  \begin{tabular}{|c|c|c|c|c|c|c|}
    \hline
  &  Initial State   & Final State & \begin{tabular}{c}First pulse\\(clock-Rydberg)\end{tabular}&  \begin{tabular}{c}Second pulse\\(clock-Rydberg)\end{tabular} & \begin{tabular}{c}Third pulse\\(ground-Rydberg)\end{tabular}&Total duration \\   
   \hline  
    SBS &$ \prod\frac{\lvert\text{c}_\uparrow\rangle + \lvert\text{c}_\downarrow\rangle}{\sqrt{2}}$ & \begin{tabular}{c} $\frac{1}{\sqrt{2}}(\lvert\text{cc}\rangle_{\text{e}} \otimes \lvert\Phi\rangle_{\text{n}}$  \\$+ \lvert\Phi\rangle_{\text{e}}  \otimes \lvert\Psi\rangle_{\text{n}} )$\end{tabular}& \begin{tabular} {cc} Rabi:& $\Omega^{(\text{S})}=1.608\Delta$\\ $\mathbb{T}_{\text{p}1}$:& $2.12\pi/\Omega^{(\text{S})}$ \end{tabular} & \begin{tabular}{cc} Rabi:&$-0.4606\Omega^{(\text{S})}$\\ $\mathbb{T}_{\text{p}2}$: & $2.85\pi/\Omega^{(\text{S})}$ \end{tabular} & \begin{tabular}{cc} Rabi:&$0.4393\Omega^{(\text{S})}$\\ $\mathbb{T}_{\text{p}3}$: & $2.71\pi/\Omega^{(\text{S})}$ \end{tabular} & $7.69\pi/\Omega^{(\text{S})}$  \\   
   \hline  
  $ \lvert  \blacktriangle \rangle$  & $\prod\frac{\lvert\text{c}_\uparrow\rangle + \lvert\text{c}_\downarrow\rangle}{\sqrt{2}}$& \begin{tabular}{c} $\frac{1}{2}[(\sqrt{3}\lvert\text{ccc}\rangle_{\text{e}} \otimes \lvert\Lambda\rangle_{\text{n}} $ \\ $  + \lvert \text{W}\rangle_{\text{e}}  \otimes \lvert \text{GHZ}\rangle_{\text{n}} )  ]$\end{tabular}
  & \begin{tabular} {cc} Rabi:& $\Omega^{(\blacktriangle)}=1.976\Delta$\\ $\mathbb{T}_{\text{p}1}^{(\blacktriangle)}$:& $5.93\pi/\Omega^{(\blacktriangle)}$ \end{tabular} & \begin{tabular}{cc} Rabi:&$-0.3735\Omega^{(\blacktriangle)}$\\ $\mathbb{T}_{\text{p}2}^{(\blacktriangle)}$: & $1.9\pi/\Omega^{(\blacktriangle)}$ \end{tabular} & \begin{tabular}{cc} Rabi:&$0.3073\Omega^{(\blacktriangle)}$\\ $\mathbb{T}_{\text{p}3}^{(\blacktriangle)}$: & $3.54\pi/\Omega^{(\blacktriangle)}$ \end{tabular} & $11.4\pi/\Omega^{(\blacktriangle)}$  \\   
   \hline          
  \end{tabular}
  \caption{Parameters for creating the two-atom super Bell state of Eq.~(\ref{Bell}) and the three-atom entangled state of Eq.~(\ref{W01}).   \label{table2}  }
\end{table*}

\section{A three-atom state including W and GHZ states}\label{Sec05}

\subsection{A three-pulse protocol}\label{Sec05A}
Here, we show a three-pulse protocol to excite the initial product state
\begin{eqnarray}
\lvert\psi(0)\rangle&=& |\text{ccc}\rangle_{\text{e}}\otimes (\lvert\uparrow\uparrow\uparrow\rangle_{\text{n}}+ \lvert\uparrow\uparrow\downarrow\rangle_{\text{n}}+ \lvert\uparrow\downarrow\uparrow\rangle_{\text{n}}
\nonumber\\
&&+\lvert\downarrow\uparrow\uparrow\rangle_{\text{n}}+\lvert \uparrow\downarrow\downarrow\rangle_{\text{n}}+ \lvert \downarrow\uparrow\downarrow\rangle_{\text{n}}+\lvert\downarrow\downarrow\uparrow\rangle_{\text{n}}
\nonumber\\
&&+  \lvert\downarrow\downarrow\downarrow\rangle_{\text{n}})/2\sqrt{2},\label{W-initialState}
\end{eqnarray}
to realize $\lvert  \blacktriangle \rangle$ in Eq.~(\ref{W01}). We will do it briefly for the pulses induce state transform similar to that in Sec.~(\ref{Sec03A}). 

In the first pulse, a UV laser with a pulse duration $\mathbb{T}_{\text{p}1}^{(\blacktriangle)}$ for the transitions 
\begin{eqnarray}
 &&\lvert \text{c}_\uparrow \rangle \xrightarrow[\text{detuned by~} \Delta]{\Omega_{\text{c}\uparrow}=\Omega^{(\blacktriangle)}}  \lvert r_+  \rangle ,\nonumber\\
 &&\lvert \text{c}_\downarrow \rangle \xrightarrow[\text{detuned by~} -\Delta]{\Omega_{\text{c}\downarrow}= \Omega^{(\blacktriangle)}}  \lvert r_-  \rangle , \nonumber
\end{eqnarray}
evolves the the input states as $\lvert \text{c}_x \text{c}_y\text{c}_z\rangle \xrightarrow{ } e^{-i\hat{H}\mathbb{T}_{\text{p}1}^{(\blacktriangle)} } \lvert \text{c}_x \text{c}_y\text{c}_z\rangle$ where $x,y,z\in\{\uparrow, \downarrow\}$. Unlike the case for creating SBS, each input state is in superposition of ground and Rydberg states after the first pulse. The second pulse for the excitation
\begin{eqnarray}
 &&\lvert \text{c}_\uparrow \rangle \xrightarrow[\text{detuned by~} \Delta]{\Omega_{\text{c}\uparrow}=\eta^{(\blacktriangle)}\Omega^{(\blacktriangle)}}  \lvert r_+  \rangle ,\nonumber\\
 &&\lvert \text{c}_\downarrow \rangle \xrightarrow[\text{detuned by~} -\Delta]{\Omega_{\text{c}\downarrow}= \eta^{(\blacktriangle)}\Omega^{(\blacktriangle)}}  \lvert r_-  \rangle , \nonumber
\end{eqnarray}
of duration $\mathbb{T}_{\text{p}2}^{(\blacktriangle)}$ leads to
\begin{eqnarray}
 e^{-i\hat{H}\mathbb{T}_{\text{p}1}^{(\blacktriangle)} }\lvert \text{c}_\uparrow \text{c}_\uparrow\text{c}_\uparrow\rangle &\xrightarrow{ }& \frac{e^{i\alpha^{(\blacktriangle)}}}{\sqrt{3}} (\lvert r_+ \text{c}_\uparrow\text{c}_\uparrow\rangle+ \lvert   \text{c}_\uparrow r_+\text{c}_\uparrow\rangle \nonumber\\
 &&+\lvert   \text{c}_\uparrow\text{c}_\uparrow r_+\rangle) ,\nonumber\\
 e^{-i\hat{H}\mathbb{T}_{\text{p}1}^{(\blacktriangle)} }\lvert \text{c}_\downarrow \text{c}_\downarrow\text{c}_\downarrow\rangle & \xrightarrow{ } &  \frac{ e^{i(\pi-\alpha^{(\blacktriangle)})}   }{\sqrt{3}} (\lvert r_- \text{c}_\downarrow\text{c}_\downarrow\rangle + \lvert \text{c}_\downarrow r_- \text{c}_\downarrow\rangle \nonumber\\
 &&+ \lvert   \text{c}_\downarrow\text{c}_\downarrow r_- \rangle ),
\label{W-pulse21}
\end{eqnarray}
and 
\begin{eqnarray}
 &&e^{-i\hat{H}\mathbb{T}_{\text{p}1}^{(\blacktriangle)} }\lvert \text{c}_x \text{c}_y\text{c}_z\rangle \xrightarrow{ } e^{i\zeta^{(\blacktriangle)} } \lvert \text{c}_x \text{c}_y\text{c}_z\rangle, 
\label{W-pulse22}
\end{eqnarray}
when one of $x,y$, and $z$ is $\downarrow$ while the other two are $\uparrow$, and
\begin{eqnarray}
 &&e^{-i\hat{H}\mathbb{T}_{\text{p}1}^{(\blacktriangle)} }\lvert \text{c}_x \text{c}_y\text{c}_z\rangle \xrightarrow{ } e^{-i\zeta^{(\blacktriangle)} } \lvert \text{c}_x \text{c}_y\text{c}_z\rangle, 
\label{W-pulse23}
\end{eqnarray}
when one of $x,y$, and $z$ is $\uparrow$ while the other two are $\downarrow$. We find that when 
\begin{eqnarray}
 |\Omega^{(\blacktriangle)}|&=& 1.976 \Delta, \nonumber\\
 \eta^{(\blacktriangle)}&=& -0.3735,\nonumber\\
 \mathbb{T}_{\text{p}1}^{(\blacktriangle)}&=& 5.933\pi/|\Omega^{(\blacktriangle)}|, \nonumber\\
 \mathbb{T}_{\text{p}2}^{(\blacktriangle)}&=& 1.902\pi/|\Omega^{(\blacktriangle)}|,   \label{W-condition-p1p2}
\end{eqnarray}
the transitions in Eqs.~(\ref{W-pulse21}),~(\ref{W-pulse22}) and~(\ref{W-pulse23}) can be realized with a fidelity over 0.995 for each input state. For infinite $V/\Delta$, $\alpha^{(\blacktriangle)}$ is $-1.036~(2.106)$~rad when $\Omega^{(\blacktriangle)}$ is positive~(negative), and $\zeta^{(\blacktriangle)}=-0.8717$~rad. After the frame transform as in Appendix~\ref{AppendixB}, the angle $\alpha^{(\blacktriangle)}$ in the state on the right side of Eq.~(\ref{W-pulse21}) becomes
\begin{eqnarray}
\alpha^{(\blacktriangle)'}&=& \alpha^{(\blacktriangle)} +\Delta(\mathbb{T}_{\text{p}1}^{(\blacktriangle)}+\mathbb{T}_{\text{p}2}^{(\blacktriangle)}), \nonumber
\end{eqnarray}  
while $\zeta^{(\blacktriangle)}$ in Eqs.~(\ref{W-pulse22}) and~(\ref{W-pulse23})  remains the same. The third pulse induces a ground-Rydberg transition as in Sec.~\ref{sec-pulse3}. We find that with the condition 
\begin{eqnarray}
 |\Omega_{\text{eff}}^{(\blacktriangle)}|&=& 0.6072\Delta\nonumber\\
  \mathbb{T}_{\text{p}3}^{(\blacktriangle)}&=&1.789\pi/\Delta, \label{p3-W}
\end{eqnarray}
the third pulse can restore the population in the states of Eq.~(\ref{W-pulse21}) back to the ground states, i.e., the states in Eq.~(\ref{W-pulse21}) evolve as
\begin{eqnarray}
 & & \frac{e^{i\alpha^{(\blacktriangle)'}}}{\sqrt{3}} (\lvert r_+ \text{c}_\uparrow\text{c}_\uparrow\rangle+ \lvert   \text{c}_\uparrow r_+\text{c}_\uparrow\rangle+\lvert   \text{c}_\uparrow\text{c}_\uparrow r_+\rangle)\nonumber\\&& \xrightarrow{ }\frac{e^{i[\alpha^{(\blacktriangle)'}+\beta^{(\blacktriangle)}]}}{\sqrt{3}} (\lvert  \text{g}_\uparrow \text{c}_\uparrow\text{c}_\uparrow\rangle+ \lvert   \text{c}_\uparrow  \text{g}_\uparrow \text{c}_\uparrow\rangle+\lvert   \text{c}_\uparrow\text{c}_\uparrow  \text{g}_\uparrow \rangle) \nonumber\\
 && =e^{i[\alpha^{(\blacktriangle)'}+\beta^{(\blacktriangle)}]}\lvert \text{W}\rangle_{\text{e}}\otimes\lvert\uparrow\uparrow\uparrow \rangle_{\text{n}} ,\nonumber\\
  &  & - \frac{ e^{-i\alpha^{(\blacktriangle)'}}   }{\sqrt{3}} (\lvert r_- \text{c}_\downarrow\text{c}_\downarrow\rangle + \lvert \text{c}_\downarrow r_- \text{c}_\downarrow\rangle + \lvert   \text{c}_\downarrow\text{c}_\downarrow r_- \rangle )\nonumber\\
 && \xrightarrow{ }- \frac{e^{i[-\alpha^{(\blacktriangle)'}-\beta^{(\blacktriangle)}]}}{\sqrt{3}} (   \lvert  \text{g}_\downarrow \text{c}_\downarrow\text{c}_\downarrow\rangle + \lvert \text{c}_\downarrow \text{g}_\downarrow \text{c}_\downarrow\rangle + \lvert   \text{c}_\downarrow\text{c}_\downarrow \text{g}_\downarrow \rangle )\nonumber\\
 && =-e^{i[-\alpha^{(\blacktriangle)'}-\beta^{(\blacktriangle)}]}\lvert \text{W}\rangle_{\text{e}}\otimes\lvert  \downarrow \downarrow \downarrow \rangle_{\text{n}} ,\nonumber
\end{eqnarray} 
where $\beta^{(\blacktriangle)}=-1.607$~($1.534$)~rad when $\Omega_{\text{eff}}^{(\blacktriangle)}$ is positive~(negative), while the states in Eqs.~(\ref{W-pulse22}) and~(\ref{W-pulse23}) remain the same because the third pulse excites the clock-Rydberg transitions. So, starting from the state of Eq.~(\ref{W-initialState}), the above pulses realize the state 
\begin{eqnarray}
&&\frac{1}{2}\Big\{\frac{\sqrt{6}}{ 2}\lvert\text{ccc}\rangle_{\text{e}} \otimes  \Big[  e^{i\zeta^{(\blacktriangle)} }\frac{\lvert \uparrow  \uparrow \downarrow  \rangle_{\text{n}} +\lvert \uparrow  \downarrow \uparrow  \rangle_{\text{n}} +\lvert\downarrow      \uparrow\uparrow  \rangle_{\text{n}}  }{\sqrt{3}}   \nonumber\\
&&+ e^{-i\zeta^{(\blacktriangle)} }\frac{\lvert  \uparrow \downarrow \downarrow \rangle_{\text{n}} +\lvert  \downarrow \uparrow \downarrow \rangle_{\text{n}} +\lvert  \downarrow \downarrow \uparrow \rangle_{\text{n}}  }{\sqrt{3}} \Big]\nonumber\\
&& +\lvert \text{W}\rangle_{\text{e}}  \otimes \frac{e^{i[\alpha^{(\blacktriangle)'}+\beta]}\lvert \uparrow  \uparrow  \uparrow    \rangle_{\text{n}} -e^{-i[\alpha^{(\blacktriangle)'}+\beta]}\lvert   \downarrow  \downarrow  \downarrow    \rangle_{\text{n}}  }{\sqrt{2}}  )  \Big\},  \nonumber
\end{eqnarray}
in which the relative phase in the nuclear-spin GHZ state can be removed by redefining the basis states. The total duration is about $11.43\pi/|\Omega^{(\blacktriangle)}|$ for creating $\lvert\blacktriangle\rangle$. A numerical test of the above protocol is shown in Fig.~\ref{W-unitary} with relevant Hamiltonians shown in Appendix~\ref{AppendixD}.

\begin{figure}
\includegraphics[width=3.2in]
{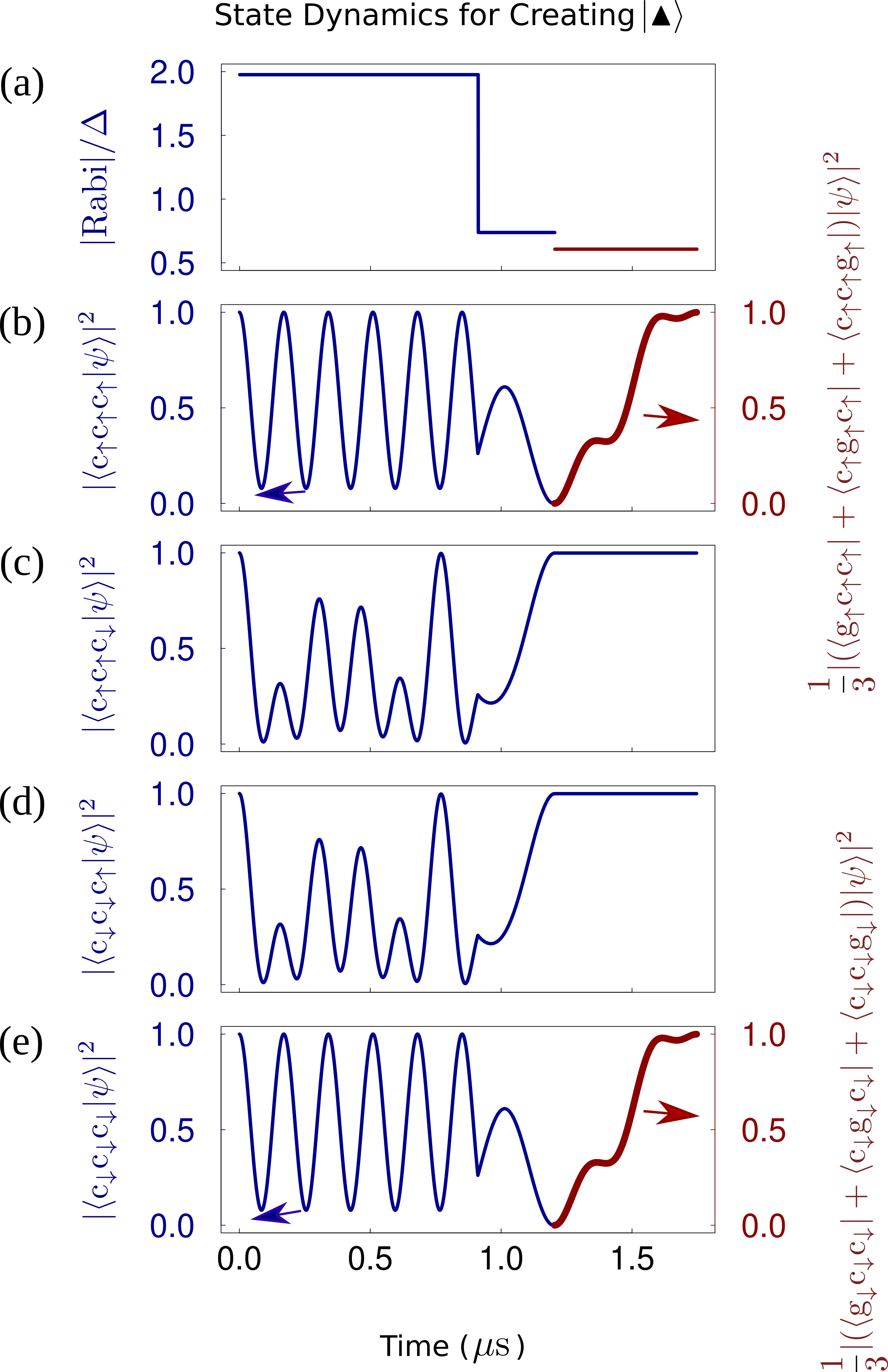}
 \caption{State dynamics of the $\lvert\blacktriangle\rangle$ protocol for different input states with $V_0/2\pi=260$~MHz and maximal Rabi frequency $\Omega^{(\blacktriangle)}/2\pi\approx3.25$~MHz when ignoring Rydberg-state decay. (a) Magnitudes of Rabi frequencies in units of $\Delta$. (b), (c), (d), and (e) show the population of the ground-state component in the wavefunction when the input states for the gate protocol are $\lvert \text{c}_\uparrow \text{c}_\uparrow \text{c}_\uparrow\rangle$, $\lvert \text{c}_\uparrow \text{c}_\uparrow\text{c}_\downarrow\rangle, \lvert \text{c}_\downarrow  \text{c}_\downarrow \text{c}_\uparrow\rangle$, and $\lvert \text{c}_\downarrow  \text{c}_\downarrow \text{c}_\downarrow\rangle$, respectively. The thin and thick curves in (b) and (e) denote population in the product and W states, respectively. For the target ground states $\lvert\text{W}\rangle_{\text{e}}  \otimes \lvert\uparrow\uparrow\uparrow\rangle_{\text{n}}$ in (b), $\lvert \text{c}_\uparrow\text{c}_\uparrow\text{c}_\downarrow \rangle$ in (c), $\lvert \text{c}_\downarrow\text{c}_\downarrow\text{c}_\uparrow\rangle$ in (d), and $\lvert\text{W}\rangle_{\text{e}}  \otimes \lvert\downarrow\downarrow\downarrow\rangle_{\text{n}}$ in (e), the final populations are $0.9986$, $0.9968$, $0.9988$, and $0.9982$ respectively, and the final phases are $\vartheta_1=\Delta\left[\mathbb{T}_{\text{p}1}^{(\blacktriangle)}+\mathbb{T}_{\text{p}2}^{(\blacktriangle)}\right]-2.532$~rad, $\vartheta_2=-0.7648$~rad, $\vartheta_3=0.9795$~rad, and $\vartheta_4=-\Delta\left[\mathbb{T}_{\text{p}1}^{(\blacktriangle)}+\mathbb{T}_{\text{p}2}^{(\blacktriangle)}\right]-0.385$~rad, respectively.  }\label{W-unitary}
\end{figure}

\begin{figure}
\includegraphics[width=2.5in]
{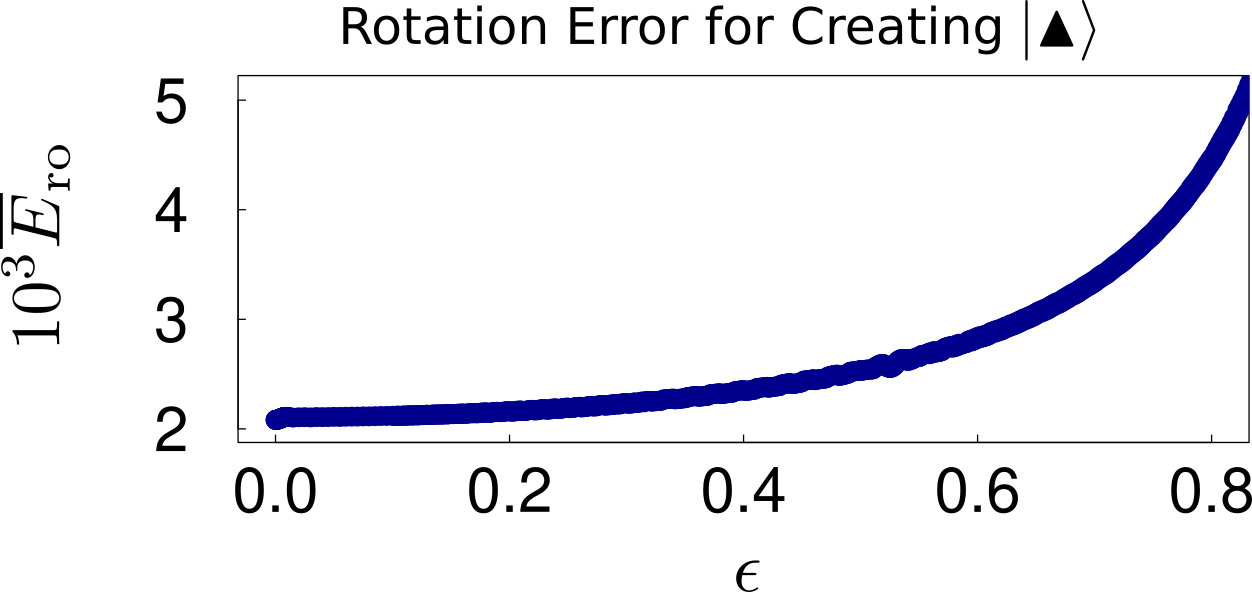}
 \caption{ Rotation error~(rescaled by $10^3$) averaged by uniformly varying the Rydberg interaction $V$ in $[(1-\epsilon)V_0,~(1+\epsilon)V_0]$ for creating $\lvert\blacktriangle\rangle$. Here, $V_0/2\pi=260$~MHz, $\Omega_{\text{eff}}/2\pi=1$~MHz, and $\Omega^{(\blacktriangle)}$ and $\Delta$ are given by Eqs.~(\ref{W-condition-p1p2}) and~(\ref{p3-W}). The fidelity to generate $\lvert\blacktriangle\rangle$ in the form of Eq.~(\ref{W-final-real}), $1- \overline{E}_{\text{ro}} -  E_{\text{decay}} $, is about $0.993$ for $\epsilon=0.8$. }\label{W-rotation-Error}
\end{figure}

\subsection{Numerical simulations}\label{Sec05B}
We consider three atoms forming a triangle in the x-y plane when the laser fields are sent along $\mathbf{z}$~\cite{Graham2019,Graham2022}, and the atoms are in a configuration where the three distances between each two atom traps are equal. To have equal laser-field strength for the three atoms, the laser is focused at the center of the triangle. When there is no fluctuation of the atomic positions, the interactions between any two Rydberg atoms are equal to the desired value $V_0$. Even so, the final state has some population loss to the target state mainly due to the finiteness of $V$, as shown in Fig.~\ref{W-unitary}.

 The Rydberg-state decay induces an error $E_{\text{decay}}\approx2.51\pi/(\tau\Delta)$, which is about $2.3\times10^{-3}$ when $\Omega_{\text{eff}}/2\pi=1$~MHz and $\tau=330~\mu$s. The fluctuation of $V$ and its finiteness result in gate errors that will be analyzed as follows. We define the state with state phases realized when $V=V_0$ as the target state 
\begin{eqnarray}
\left\lvert\vee\right\rangle&=&\frac{1}{2}\Big\{\frac{\sqrt{6}}{ 2}\lvert\text{ccc}\rangle_{\text{e}} \otimes  \Big[  e^{i\vartheta_1 }\frac{\lvert \uparrow  \uparrow \downarrow  \rangle_{\text{n}} +\lvert \uparrow  \downarrow \uparrow  \rangle_{\text{n}} +\lvert\downarrow      \uparrow\uparrow  \rangle_{\text{n}}  }{\sqrt{3}}   \nonumber\\
&&+ e^{i\vartheta_2 }\frac{\lvert  \uparrow \downarrow \downarrow \rangle_{\text{n}} +\lvert  \downarrow \uparrow \downarrow \rangle_{\text{n}} +\lvert  \downarrow \downarrow \uparrow \rangle_{\text{n}}  }{\sqrt{3}} \Big]\nonumber\\
&& +\lvert \text{W}\rangle_{\text{e}}  \otimes \frac{e^{i\vartheta_3 }\lvert \uparrow  \uparrow  \uparrow    \rangle_{\text{n}} +e^{i\vartheta_4 }\lvert   \downarrow  \downarrow  \downarrow    \rangle_{\text{n}}  }{\sqrt{2}}   \Big\},  \label{W-final-real}
\end{eqnarray}
where the four angles $\vartheta_{k}, k=1\textendash4$ are given in the caption of Fig.~\ref{W-unitary}. If we define new nuclear-spin states by absorbing phases to them as
\begin{eqnarray}
 \text{new} \lvert \text{g} \rangle_{\text{e}}&=&   e^{i(2\vartheta_4+7\vartheta_3-6\vartheta_1)/9} \lvert \text{g}   \rangle _{\text{e}}  ,\nonumber\\
\text{new} \lvert \text{c} \rangle_{\text{e}}&=&  e^{i(3\vartheta_1+\vartheta_3-\vartheta_4)/9} \lvert \text{c}   \rangle _{\text{e}},\nonumber\\
 \text{new} \lvert \downarrow  \rangle _{\text{n}}&=&  e^{i(\vartheta_4-\vartheta_3)/3} \lvert \downarrow  \rangle _{\text{n}} , \label{basisDef-W}
\end{eqnarray}
the four phases $\vartheta_{1}\textendash\vartheta_{4}$ of the state in Eq.~(\ref{W-final-real}) become $0, \vartheta_2-\vartheta_1+(\vartheta_3-\vartheta_4)/3$, and $0$, respectively. The basis transform is not necessary since in real experiments the interesting part of the state is the electrons-nuclei entanglement. So, the state in Eq.~(\ref{W-final-real}) is the target $\lvert\blacktriangle\rangle$ state. To quantify the error from the fluctuation of $V$, we define $E_{\text{ro}} =1-\langle \vee |\hat{\rho}|\vee \rangle$ as the rotation error, where $\hat{\rho}$ is the density matrix of the actual state realized with the three-pulse protocol. With unitary dynamics, the rotation error becomes $1-\lvert\langle  \psi\lvert\vee \rangle |^2$ where $\lvert \psi\rangle$ is the final state evolved from Eq.~(\ref{initialState}). To evaluate the rotation error, we consider the average $ \overline{E}_{\text{ro}} = \int E_{\text{ro}}(V) dV/\int dV$, where the integration is over $V\in[(1-\epsilon)V_0,~(1+\epsilon)V_0]$. With a uniform distribution of $V$, the numerical results for different $\epsilon$ are shown in Fig.~\ref{W-rotation-Error}. The average rotation error $\overline{E}_{\text{ro}} $ increases from $2$ to $4.5\times10^{-3}$ when $\epsilon$ grows from $0$ to $0.8$. The fidelity can be about 0.993 even if $\epsilon$ is as large as $0.8$.  

Note that the numerical example shown above can be optimized so as to have a faster creation of the state. For example, Appendix~\ref{AppendixE} shows another set of parameters to create $\lvert\blacktriangle\rangle$ with a faster speed. The purpose of this paper is to reveal the possibility to create exotic multi-atom entanglement between electrons and nuclear spins.

\section{Creating lower-dimensional entanglement by measurement}\label{Sec06}
SBS and $\lvert\blacktriangle\rangle$ are highly entangled. Here, we show that it is possible to start from them to create other types of entangled states. Taking SBS as an example, SBS incorporates three small Bell states as discussed in Sec.~\ref{Sec03} and it is a Bell-like state by itself, shown in Eq.~(\ref{Bell4}). This means that it is in principle possible to extract the respective ``smaller'' Bell state. 

\subsection{Measurement of one atom at a time in SBS}
Below, we show that by measuring one or two atoms, the SBS in Eq.~(\ref{SBS-final}) can be projected to different entangled states depending on the measurement results. In other words, it is possible to create lower-dimensional entanglement by measurement of SBS. 

By measuring one of the two atoms, one can find that the atom is in the ground state or not, by detecting the light scattered due to the transition between the ground state and $(6s6p)^1P_1)$~\cite{Jenkins2022}. However, we need to ensure that the measurement does not change the nuclear-spin states which need efforts to quench the hyperfine-interaction-induced spin mixing~\cite{Reichenbach2007}. If the first atom is measured, and the result is that no light is collected, the state in Eq.~(\ref{SBS-final}) becomes
\begin{eqnarray}
 \lvert \xi_1\rangle&=&\frac{1}{\sqrt{6}}\Big[ \lvert \text{cg}\rangle_{\text{e}} \otimes\left(e^{i\theta_1}\lvert \uparrow\uparrow\rangle_{\text{n}}+ e^{i\theta_2} \lvert \downarrow\downarrow\rangle_{\text{n}}\right)\nonumber\\
&& + e^{i\theta_3}\sqrt{2}\lvert \text{cc}\rangle_{\text{e}} \otimes\left(\lvert \uparrow\downarrow\rangle_{\text{n}} +\lvert \downarrow\uparrow\rangle_{\text{n}}\right) \Big].\label{meansure01}
\end{eqnarray}
Then, measuring the second atom in the state of Eq.~(\ref{meansure01}) has two outcomes. If no light is detected, the state in Eq.~(\ref{meansure01}) collapses to
\begin{eqnarray}
 \lvert \xi_{1a}\rangle&=& \frac{1}{\sqrt{2}}\lvert \text{cc}\rangle_{\text{e}} \otimes\left(\lvert \uparrow\downarrow\rangle_{\text{n}} +\lvert \downarrow\uparrow\rangle_{\text{n}}\right) ,\label{meansure02}
\end{eqnarray}
but if light is detected, the state in Eq.~(\ref{meansure01}) collapses to
\begin{eqnarray}
 \lvert \xi_{1b}\rangle&=& \frac{1}{\sqrt{2}} \lvert \text{cg}\rangle_{\text{e}} \otimes\left(e^{i\theta_1}\lvert \uparrow\uparrow\rangle_{\text{n}}+ e^{i\theta_2} \lvert \downarrow\downarrow\rangle_{\text{n}}\right).\label{meansure03}
\end{eqnarray}
When the measurement of the first atom results in light detected, meaning that the atom is in $\lvert\text{g}\rangle$, the state in Eq.~(\ref{SBS-final}) becomes
\begin{eqnarray}
 \lvert \xi_{2}\rangle&=& \frac{1}{\sqrt{2}}\lvert \text{gc}\rangle_{\text{e}} \otimes  \left(e^{i\theta_1}\lvert \uparrow\uparrow\rangle_{\text{n}}+ e^{i\theta_2} \lvert \downarrow\downarrow\rangle_{\text{n}}\right).\label{meansure04}
\end{eqnarray}
The state in Eq.~(\ref{meansure02}) is an entangled nuclear-spin state in the clock state. The states in Eqs.~(\ref{meansure03}) and~(\ref{meansure04}) can be converted to 
\begin{eqnarray}
 \lvert \xi_{1b}'\rangle&=&  \lvert \text{gg}\rangle_{\text{e}} \otimes\left(e^{i\theta_1}\lvert \uparrow\uparrow\rangle_{\text{n}}+ e^{i\theta_2} \lvert \downarrow\downarrow\rangle_{\text{n}}\right),\nonumber
\end{eqnarray}
and
\begin{eqnarray}
 \lvert \xi_{2}'\rangle&=& \lvert \text{cc}\rangle_{\text{e}} \otimes  \left(e^{i\theta_1}\lvert \uparrow\uparrow\rangle_{\text{n}}+ e^{i\theta_2} \lvert \downarrow\downarrow\rangle_{\text{n}}\right), \nonumber
\end{eqnarray}
respectively by using a $\pi$ pulse of ground-clock transition in the first atom~(with an extra phase determined by the Rabi frequency), see texts around Eq.~(\ref{g-c-transition}).

The above discussions are based on detecting ground-state atoms without losing the atoms. It was reported in Ref.~\cite{Covey2019} that high-fidelity~(over 0.9999) measurement of strontium atoms can proceed with a survival probability over 0.999. Imaging AEL atoms with nuclear spins would require extra efforts. Recently, Ref.~\cite{Jenkins2022} demonstrated imaging of the ground-state $^{171}$Yb atoms in tweezers with a fidelity about $0.997$ and a survival probability around $2\%\textendash3\%$~(near-unit fidelity can be achieved with larger atom loss probability). 

\subsection{Measurement of two atoms at a time in SBS}
Measurement of two atoms can also lead to interesting states. For example, if no light is detected when measuring the two atoms initially in the state of Eq.~(\ref{SBS-final}), the state collapses to  
\begin{eqnarray}
 \lvert \Xi_1 \rangle&=&\frac{1}{\sqrt{2}} \lvert \text{cc}\rangle_{\text{e}} \otimes\left(\lvert \uparrow\downarrow\rangle_{\text{n}} +\lvert \downarrow\uparrow\rangle_{\text{n}}\right).\nonumber
\end{eqnarray}
But if light is detected, and if it is possible to collect light scattered from the two nearby atoms when neither any device nor the environment know which atom scatters the light, the state collapses to 
\begin{eqnarray}
 \lvert \Xi_2 \rangle&=&\frac{1}{2} (\lvert \text{cg}\rangle_{\text{e}}+\lvert  \text{gc}_{\text{e}}\rangle ) \otimes (e^{i\theta_1}\lvert \uparrow\uparrow\rangle_{\text{n}}+ e^{i\theta_2} \lvert \downarrow\downarrow\rangle_{\text{n}} ),\nonumber
\end{eqnarray}
which is a hyperentangled state where both the electronic states and the nuclear-spin states of the two atoms are maximally entangled~\cite{Shi2021pra}. Here, we need to emphasize that though it was not tested with AEL atoms for collecting light scattered from two atoms without distinguishing the light pathway, the physical principle is similar to the double-split interference experiment.

\subsection{Measurement of $\lvert\blacktriangle\rangle$}
The three-atom state $\lvert\blacktriangle\rangle$ in the form of Eq.~(\ref{W-final-real}) can also be used for creating familiar interesting entangled states. Because of the descriptions shown above about SBS, we only briefly discuss the measurement of $\lvert\blacktriangle\rangle$. 

We first discuss the outcome of measuring one atom. If we measure the first atom and find light scattered, then Eq.~(\ref{W-final-real}) collapses to $\lvert\text{gcc}\rangle_{\text{e}}  \otimes (e^{i\vartheta_3 }\lvert \uparrow  \uparrow  \uparrow    \rangle_{\text{n}} +e^{i\vartheta_4 }\lvert   \downarrow  \downarrow  \downarrow    \rangle_{\text{n}} )/\sqrt{2} $. This state can be further converted to $\lvert\text{ccc}\rangle_{\text{e}}  \otimes (e^{i\vartheta_3 }\lvert \uparrow  \uparrow  \uparrow    \rangle_{\text{n}} +e^{i\vartheta_4 }\lvert   \downarrow  \downarrow  \downarrow    \rangle_{\text{n}} )/\sqrt{2} $ via the ground-clock transition in the first atom, which is the canonical GHZ state considering that the relative phase in it can be absorbed by defining new spin basis states as in Eq.~(\ref{basisDef-W}). But if no light is scattered when measuring the first atom, the state collapses to  
\begin{eqnarray}
&&\frac{1}{2}\Big\{\frac{\sqrt{6}}{ 2}\lvert\text{ccc}\rangle_{\text{e}} \otimes  \Big[  e^{i\vartheta_1 }\frac{\lvert \uparrow  \uparrow \downarrow  \rangle_{\text{n}} +\lvert \uparrow  \downarrow \uparrow  \rangle_{\text{n}} +\lvert\downarrow      \uparrow\uparrow  \rangle_{\text{n}}  }{\sqrt{3}}   \nonumber\\
&&+ e^{i\vartheta_2 }\frac{\lvert  \uparrow \downarrow \downarrow \rangle_{\text{n}} +\lvert  \downarrow \uparrow \downarrow \rangle_{\text{n}} +\lvert  \downarrow \downarrow \uparrow \rangle_{\text{n}}  }{\sqrt{3}} \Big]\nonumber\\
&& +\frac{1}{\sqrt{2}}(\lvert \text{cgc}\rangle_{\text{e}} +\lvert \text{ccg}\rangle_{\text{e}} ) \otimes \frac{e^{i\vartheta_3 }\lvert \uparrow  \uparrow  \uparrow    \rangle_{\text{n}} +e^{i\vartheta_4 }\lvert   \downarrow  \downarrow  \downarrow    \rangle_{\text{n}}  }{\sqrt{2}}     \Big\}.  \nonumber\\
\label{W-measure-1}
\end{eqnarray}
Measurement of the second atom in the state of Eq.~(\ref{W-measure-1}) can lead to two outcomes. (i) If no light is detected, then Eq.~(\ref{W-measure-1}) collapses to the state where $\frac{1}{\sqrt{2}}(\lvert \text{cgc}\rangle_{\text{e}} +\lvert \text{ccg}\rangle_{\text{e}} )$ in the last line of Eq.~(\ref{W-measure-1}) is replaced by $\lvert \text{ccg}\rangle_{\text{e}}$, and if we continue to measure the third atom, and detect light, then the state finally collapses to $\lvert\text{ccg}\rangle_{\text{e}}  \otimes (e^{i\vartheta_3 }\lvert \uparrow  \uparrow  \uparrow    \rangle_{\text{n}} +e^{i\vartheta_4 }\lvert   \downarrow  \downarrow  \downarrow    \rangle_{\text{n}} )/\sqrt{2} $. (ii) If light is detected, Eq.~(\ref{W-measure-1}) collapses to $\lvert\text{cgc}\rangle_{\text{e}}  \otimes (e^{i\vartheta_3 }\lvert \uparrow  \uparrow  \uparrow    \rangle_{\text{n}} +e^{i\vartheta_4 }\lvert   \downarrow  \downarrow  \downarrow    \rangle_{\text{n}} )/\sqrt{2} $, which can be converted to the canonical GHZ state as discussed above. 

We then discuss measuring three atoms at a time. We consider the case that we can collect light scattered from the atoms when neither any device nor the environment know which atom scatters the light. If we find light scattered, Eq.~(\ref{W-final-real}) collapses to
\begin{eqnarray}
 \frac{1}{\sqrt{2}} \lvert \text{W}\rangle_{\text{e}}  \otimes \left(e^{i\vartheta_3 }\lvert \uparrow  \uparrow  \uparrow    \rangle_{\text{n}} +e^{i\vartheta_4 }\lvert   \downarrow  \downarrow  \downarrow    \rangle_{\text{n}} \right) ,  \nonumber
\end{eqnarray}
which is a hyerentangled state where the electronic state of the three atoms is in the maximally entangled W state, and the nuclear-spin state is in the maximally entangled GHZ state. In principle, it is possible to convert the electronic W state to the electronic GHZ state~\cite{PhysRevA.106.052613} so that the electronic GHZ state and the nuclear-spin GHZ state can simultaneously exist in three atoms. 

\section{discussions}\label{Sec07}
The speed of the nuclear-spin quantum gates is large. To discuss the speed of a Rydberg-mediated quantum gate, it is useful to denote the gate duration in terms of $\pi/\Omega_{\text{m}}$ where $\Omega_{\text{m}}$ is the maximal Rabi frequency during the protocol. To our knowledge, the fastest Rydberg-blockade-based entangling gate that can be transferred to the CZ gate via single-qubit gates needed two laser pulses of total duration about $2.732\pi/\Omega_{\text{m}}$~\cite{Levine2019}. The gate of~\cite{Levine2019} depends on exciting only one of the two qubit states to Rydberg states, which means that it fits qubits defined either by the hyperfine substates as used in the experiments of~\cite{Levine2019,Bluvstein2022,Graham2022} or by the ground-clock states~(for which an experiment was done~\cite{Schine2022} but with another gate protocol~\cite{Mitra2020}). Of course, one can try to use it for nuclear spin qubits as experimentally done in Ref.~\cite{Ma2022} with a $\Omega_{\text{m}}$ small compared to the the Zeeman splitting between Rydberg states. The small $\Omega_{\text{m}}$ results in relatively long gate durations, which was possibly why the gate fidelity of Ref.~\cite{Ma2022} was even smaller than that of Ref.~\cite{Levine2019} though in principle AEL atoms can allow high-fidelity entanglement generation due to the advantages possessed by them as introduced in Sec.~\ref{sec01}. In this paper, the two or three-pulse gate has a total duration $2.589\pi/\Omega_{\text{m}}$ or $5.054\pi/\Omega_{\text{m}}$. With $\Omega_{\text{m}}/2\pi=13$~MHz realizable~\cite{Madjarov2020}, even the slower three-pulse gate in this paper would have a short duration $0.19~\mu$s. This gives hope to realize fast, high-fidelity nuclear-spin entanglement.

The theory shown in this paper is based on using a magnetic field below $10$~G for specifying the quantization axis of the atoms. The numerical simulations for the nuclear-spin quantum gates shown in Figs.~\ref{Gate-unitary},~\ref{Gate-rotation},~\ref{Gate-2pulses-unitary}, and~\ref{Gate-rotation-2p} assume $2\Delta/2\pi\approx4$~MHz, which corresponds to a B-field of about $2$~G. For the data of SBS shown in Figs.~\ref{SBS-unitary} and~\ref{SBS-rotation}, the Zeeman splitting is $2\Delta/2\pi\approx2.8$~MHz which corresponds to $B\approx1.5$~G. For the data about $\lvert\blacktriangle\rangle$ shown in Figs.~\ref{W-unitary} and~\ref{W-rotation-Error}, the field $B\approx1.7$~G is assumed. If Rabi frequencies used in them are increased by three times which can lead to much faster entanglement generations with higher fidelities, B-fields below 10~G are still sufficient. So, the protocols in this paper are compatible with recent experimental facilities studying nuclear spins, where a B-field of $4.11$~G~\cite{Ma2022}, $11$~G~\cite{Barnes2022}, or a value in the range $(0,~4.9]$~G~\cite{Jenkins2022} was used with $^{87}$Sr~\cite{Barnes2022} or $^{171}$Yb~\cite{Ma2022,Jenkins2022}. These bring hope to realize functional quantum devices with nuclear-spin quantum memories~\cite{Daley2008,Gorshkov2009,Omanakuttan2021,Shi2021pra,Chen2022,Wu2022}.

The numerical simulations in this paper are performed with $^{171}$Yb for it can simplify the analysis due to that it has the simplest nuclear spin $I=1/2$. Extending the theory to isotopes with larger $I$ is possible. For example, for nuclear-spin qubits defined with $1-I$ and $-I$ in $^{87}$Sr, the more difficult ground-Rydberg transition was demonstrated~\cite{Barnes2022}, and the clock-Rydberg transition was demonstrated with $^{88}$Sr~\cite{Madjarov2020}. Moreover, when the Rydberg Rabi frequencies for the two nuclear spin states do not have the same magnitude as in $^{171}$Yb shown in Eq.~(\ref{RabiRelation}), application of the theory in this paper may lead to more exotic entanglement because different Rydberg excitation rates in the two qubit states of each atom can result in different dynamics useful for quantum control~\cite{Shi2020}.

\section{conclusions}\label{Sec08}
We present fast and high-fidelity nuclear-spin controlled-phase gates of an arbitrary phase in AEL atoms. By exciting two AEL atoms with global laser fields where both nuclear spin qubit states of each atom are Rydberg excited, a controlled-phase gate of an arbitrary phase can be realized via three laser pulses of total duration $5.05\pi/\Omega_{\text{m}}$, where $\Omega_{\text{m}}$ is the largest Rydberg Rabi frequency during the pulses. We also show that if spin-dependent Stark shift~($\sim\Omega_{\text{m}}$) is realizable, the same gate can be realized with two laser pulses of total duration $2.59\pi/\Omega_{\text{m}}$. With large $\Omega_{\text{m}}$ for the transition between the clock and Rydberg states of AEL realizable as recently demonstrated in experiments, and concerning that fast gates are easier to atain a higher fidelity when executed, the protocols can enable high-fidelity entanglement in nuclear spins of AEL atoms.   

We also present two protocols to entangle electrons and nuclear spins in AEL. First, we show a super Bell state~(SBS) entangled between the electrons and nuclei of two atoms. SBS is like a large ``Bell'' state entangled between three ``small'' Bell states and one product state. For three atoms, we show a protocol to realize a highly entangled state that includes W and GHZ states simultaneously. These two multi-atom entangling protocols have durations $7.69\pi/\Omega_{\text{m}}$ and $11.4\pi/\Omega_{\text{m}}$, respectively, and can be used for preparing Bell, hyperentangled, and nuclear-spin GHZ states based on measurement.

\section*{ACKNOWLEDGMENTS}
The author thanks Yan Lu for useful discussions. This work is supported by the National Natural Science Foundation of China under Grants No. 12074300 and No. 11805146, the Innovation Program for Quantum Science and Technology 2021ZD0302100, and the Fundamental Research Funds for the Central Universities.

\appendix{}

\section{Hamiltonians for generating the nuclear-spin gate}\label{AppendixA}
With $\pi$ polarized laser fields, the projection of the atomic angular momentum is conserved, so that the dynamics of the atomic states can be specified by the total $m_F$ of the initial two-atom state. For $\lvert \text{c}_\uparrow \text{c}_\uparrow\rangle\equiv|\text{cc}\rangle_{\text{e}}\otimes \lvert\uparrow\uparrow\rangle_{\text{n}}$, the Hamiltonian is
\begin{eqnarray}
 \hat{H}_{\uparrow\uparrow} &=& \left(\begin{array}{cccc}
 V+2\Delta& \frac{\Omega_{\text{c}\uparrow}}{2}& \frac{\Omega_{\text{c}\uparrow}}{2}& 0 \\
 \frac{\Omega_{\text{c}\uparrow}^\ast}{2}& \Delta&  0&  \frac{\Omega_{\text{c}\uparrow}}{2}\\
 \frac{\Omega_{\text{c}\uparrow}^\ast}{2}& 0& \Delta&  \frac{\Omega_{\text{c}\uparrow}}{2}\\
 0&  \frac{\Omega_{\text{c}\uparrow}^\ast}{2}& \frac{\Omega_{\text{c}\uparrow}^\ast}{2} &  0
             \end{array}\right)\label{Huu}
\end{eqnarray}
in the basis of
\begin{eqnarray}
 \{\lvert r_+r_+\rangle, \lvert r_+\text{c}_\uparrow\rangle, \lvert \text{c}_\uparrow r_+\rangle, \lvert \text{c}_\uparrow \text{c}_\uparrow\rangle \}.\nonumber
\end{eqnarray}
 For $\lvert \text{c}_\downarrow \text{c}_\downarrow\rangle\equiv|\text{cc}\rangle_{\text{e}}\otimes \lvert\downarrow\downarrow\rangle_{\text{n}}$, the Hamiltonian is
\begin{eqnarray}
 \hat{H}_{\downarrow\downarrow} &=& \left(\begin{array}{cccc}
 V-2\Delta& \frac{\Omega_{\text{c}\downarrow}}{2}& \frac{\Omega_{\text{c}\downarrow}}{2}& 0 \\
 \frac{\Omega_{\text{c}\downarrow}^\ast}{2}&- \Delta&  0&  \frac{\Omega_{\text{c}\downarrow}}{2}\\
 \frac{\Omega_{\text{c}\downarrow}^\ast}{2}& 0&- \Delta&  \frac{\Omega_{\text{c}\downarrow}}{2}\\
 0&  \frac{\Omega_{\text{c}\downarrow}^\ast}{2}& \frac{\Omega_{\text{c}\downarrow}^\ast}{2} &  0
             \end{array}\right)  \label{Hdd}
\end{eqnarray}
in the basis of
\begin{eqnarray}
 \{\lvert r_-r_-\rangle, \lvert r_-\text{c}_\downarrow\rangle, \lvert \text{c}_\downarrow r_-\rangle,  \lvert \text{c}_\downarrow \text{c}_\downarrow\rangle \}.\nonumber
\end{eqnarray}
 For $\lvert \text{c}_\uparrow \text{c}_\downarrow\rangle\equiv|\text{cc}\rangle_{\text{e}}\otimes \lvert\uparrow\downarrow\rangle_{\text{n}}$, the Hamiltonian is
\begin{eqnarray}
 \hat{H}_{\uparrow\downarrow} &=&  \left(\begin{array}{cccc}
 V & \frac{\Omega_{\text{c}\downarrow}}{2}& \frac{\Omega_{\text{c}\uparrow}}{2}& 0 \\
 \frac{\Omega_{\text{c}\downarrow}^\ast}{2}& \Delta&  0&  \frac{\Omega_{\text{c}\uparrow}}{2}\\
 \frac{\Omega_{\text{c}\uparrow}^\ast}{2}& 0& -\Delta&  \frac{\Omega_{\text{c}\downarrow}}{2}\\
 0&  \frac{\Omega_{\text{c}\uparrow}^\ast}{2}& \frac{\Omega_{\text{c}\downarrow}^\ast}{2} &  0
             \end{array}\right)    \label{Hud}
\end{eqnarray}
in the basis of
\begin{eqnarray}
 \{\lvert r_+r_-\rangle, \lvert r_+\text{c}_\downarrow\rangle, \lvert \text{c}_\uparrow r_-\rangle,   \lvert \text{c}_\uparrow \text{c}_\downarrow\rangle \},\nonumber
\end{eqnarray}
while for $\lvert  \text{c}_\downarrow \text{c}_\uparrow\rangle\equiv|\text{cc}\rangle_{\text{e}}\otimes \lvert\downarrow\uparrow\rangle_{\text{n}}$, the Hamiltonian is similar to Eq.~(\ref{Hud}).

\section{Frame transform}\label{FrameTrans}\label{AppendixB}
Here, we show the details about a frame transform in the analyses for creating SBS. The description from Sec.~\ref{sec-pulse1} to Sec.~\ref{sec-pulse2} is in a rotating frame that transfers the Hamiltonian $\hat{\mathbb{H}}$ to $e^{it\hat{R}}\hat{\mathbb{H}} e^{-it\hat{R}} - \hat{R}\equiv \hat{H}$ with
\begin{eqnarray}
 \hat{R} &=& \omega (\lvert r_+ \rangle\langle r_+|+ \lvert r_-  \rangle\langle r_-| )-\omega_{\text{gc}}(\lvert \text{g}_\uparrow \rangle\langle \text{g}_\uparrow|+\lvert \text{g}_\downarrow \rangle\langle \text{g}_\downarrow|), \nonumber
\end{eqnarray}
where $\omega\pm\Delta=E_\pm$ is the energy separation~(divided by the reduced Planck constant) between the clock state and $\lvert r_\pm\rangle$, and $\omega_{\text{gc}}$ is the energy separation between the ground and clock states; the energy is measured in reference to that of the clock state, and we ignore the energy separation between $\lvert \text{c}_\uparrow\rangle$ and $\lvert \text{c}_\downarrow\rangle$ and that between $\lvert \text{g}_\uparrow\rangle$ and $\lvert \text{g}_\downarrow\rangle$ since they are orders of magnitude smaller than the Rabi frequencies considered in this paper due to the tiny nuclear-spin Zeeman splitting in a Gauss-scale magnetic field. In the third pulse shown below, another rotating frame defined with 
\begin{eqnarray}
 \hat{R}' &=& E_+\lvert r_+ \rangle\langle r_+|+ E_-\lvert r_-  \rangle\langle r_-| ) -\omega_{\text{gc}}(\lvert \text{g}_\uparrow \rangle\langle \text{g}_\uparrow|+\lvert \text{g}_\downarrow \rangle\langle \text{g}_\downarrow\rvert) \nonumber
\end{eqnarray}
is used. The frame transform from $\hat{R}$ to $\hat{R}'$ changes the phase $\alpha$ in the first two equations of Eq.~(\ref{pulse2}) to
\begin{eqnarray}
\alpha'&=&\alpha+\Delta\left[\mathbb{T}_{\text{p}1}^{(\text{\tiny{S}})}+\mathbb{T}_{\text{p}2}^{(\text{\tiny{S}})}\right].  \nonumber
\end{eqnarray}

\section{Hamiltonians in the third pulses for creating SBS}\label{AppendixC}
The third pulse for creating SBS described in Sec.~\ref{sec-pulse3} and for the third pulse for creating $\lvert \blacktriangle\rangle$ are similar, both by two-photon Rydberg laser excitation of the transitions
\begin{eqnarray}
 &&\lvert \text{g}_\uparrow \rangle \xrightarrow[]{\Omega_{\text{gr}\uparrow} = \Omega_{\text{eff}}(1+e^{2it\Delta})}  \lvert r_+  \rangle ,\nonumber\\
 &&\lvert \text{g}_\downarrow \rangle \xrightarrow[]{\Omega_{\text{cr}\downarrow} = -\Omega_{\text{eff}}(1+e^{-2it\Delta})}  \lvert r_-  \rangle , \nonumber\label{ground-Rydberg02}
\end{eqnarray}
where the sign difference between the ground-Rydberg Rabi frequencies for the $\uparrow$ and $\downarrow$ states is from the angular momentum selection rules~\cite{Shi2021}. Though the Hamiltonians are simple, we would like to show them without using the entangled basis states as in Sec.~\ref{sec-pulse3}. We show the simple Hamiltonians also because we would like to clarify that the Hamiltonians are in a rotating frame defined by the energy of the atomic states, which is why we need the frame transform described in Sec.~\ref{FrameTrans} when moving from the discussion of the second pulse to the discussion of the third pulse.

For the input state $\lvert \text{c}_\uparrow \text{c}_\uparrow\rangle$ which is  $e^{i\alpha'}\frac{\lvert r_+ \text{c}_\uparrow\rangle+\lvert \text{c}_\uparrow r_+\rangle}{\sqrt{2}}$ at the beginning of the third pulse, the Hamiltonian is
\begin{eqnarray}
 \hat{H}_{\uparrow\uparrow} &=& \left(\begin{array}{cccc}
0& 0& \frac{\Omega_{\text{g}\uparrow}}{2}& 0 \\
 0& 0&  0&  \frac{\Omega_{\text{g}\uparrow}}{2}\\
 \frac{\Omega_{\text{g}\uparrow}^\ast}{2}& 0& 0&  0\\
 0&  \frac{\Omega_{\text{g}\uparrow}^\ast}{2}&  0&  0
             \end{array}\right)\label{H-third-uu}
\end{eqnarray}
in the basis of
\begin{eqnarray}
 \{ \lvert r_+\text{c}_\uparrow\rangle, \lvert \text{c}_\uparrow r_+\rangle, \lvert \text{g}_\uparrow \text{c}_\uparrow\rangle, \lvert \text{c}_\uparrow \text{g}_\uparrow\rangle \}.\nonumber
\end{eqnarray}
For the input state $\lvert \text{c}_\downarrow \text{c}_\downarrow\rangle$ which is  $e^{i(\pi-\alpha')}\frac{\lvert r_- \text{c}_\downarrow\rangle +\lvert \text{c}_\downarrow r_-\rangle}{\sqrt{2}}$ at the beginning of the third pulse, the Hamiltonian is
\begin{eqnarray}
 \hat{H}_{\downarrow\downarrow} &=& \left(\begin{array}{cccc}
 0& 0& \frac{\Omega_{\text{g}\downarrow}}{2}& 0 \\
 0& 0&  0&  \frac{\Omega_{\text{g}\downarrow}}{2}\\
 \frac{\Omega_{\text{g}\downarrow}^\ast}{2}& 0& 0&  0\\
 0&  \frac{\Omega_{\text{g}\downarrow}^\ast}{2}&  0&  0
             \end{array}\right)\label{H-third-dd}
\end{eqnarray}
in the basis of
\begin{eqnarray}
 \{ \lvert r_-\text{c}_\downarrow\rangle, \lvert \text{c}_\downarrow r_-\rangle, \lvert \text{g}_\downarrow \text{c}_\downarrow\rangle, \lvert \text{c}_\downarrow \text{g}_\downarrow\rangle \}.\nonumber
\end{eqnarray}

\section{Hamiltonians for creating $\lvert \blacktriangle\rangle$}\label{AppendixD}
For the ground-Rydberg transition in the third pulse of the protocol for generating $\lvert \blacktriangle\rangle$, the dynamics is similar to that in Sec.~\ref{AppendixB}. Below, we show Hamiltonians for the first and second pulses. For $\lvert \text{c}_\uparrow\text{c}_\uparrow \text{c}_\uparrow\rangle\equiv|\text{ccc}\rangle_{\text{e}}\otimes \lvert\uparrow\uparrow\uparrow\rangle_{\text{n}}$, the Hamiltonian is
 \begin{widetext}
\begin{eqnarray}
   && \left(\begin{array}{cccccccc}
 \tiny{ 3V+3\Delta}& \frac{\Omega_{\text{c}\uparrow}}{2}& \frac{\Omega_{\text{c}\uparrow}}{2}& \frac{\Omega_{\text{c}\uparrow}}{2}& 0& 0& 0& 0 \\
\frac{\Omega_{\text{c}\uparrow}^\ast}{2}&  V+2\Delta& 0&  0& \frac{\Omega_{\text{c}\uparrow}}{2}&  \frac{\Omega_{\text{c}\uparrow}}{2}&   0&  0\\
\frac{\Omega_{\text{c}\uparrow}^\ast}{2}& 0&  V+2\Delta&  0& \frac{\Omega_{\text{c}\uparrow}}{2}&  0&  \frac{\Omega_{\text{c}\uparrow}}{2}&   0\\
\frac{\Omega_{\text{c}\uparrow}^\ast}{2}& 0& 0& V+2\Delta & 0 &    \frac{\Omega_{\text{c}\uparrow}}{2}&  \frac{\Omega_{\text{c}\uparrow}}{2}&   0\\
0&  \frac{\Omega_{\text{c}\uparrow}^\ast}{2}&  \frac{\Omega_{\text{c}\uparrow}^\ast}{2}& 0& \Delta&  0&   0&  \frac{\Omega_{\text{c}\uparrow}}{2}\\
0&  \frac{\Omega_{\text{c}\uparrow}^\ast}{2}& 0&  \frac{\Omega_{\text{c}\uparrow}^\ast}{2}& 0 & \Delta&  0&  \frac{\Omega_{\text{c}\uparrow}}{2}\\
0& 0&  \frac{\Omega_{\text{c}\uparrow}^\ast}{2}&    \frac{\Omega_{\text{c}\uparrow}^\ast}{2}&0& 0& \Delta&  \frac{\Omega_{\text{c}\uparrow}}{2}\\
0& 0& 0& 0&  \frac{\Omega_{\text{c}\uparrow}^\ast}{2}&  \frac{\Omega_{\text{c}\uparrow}^\ast}{2}& \frac{\Omega_{\text{c}\uparrow}^\ast}{2} &  0
             \end{array}\right)\label{Huuu}
\end{eqnarray}
\end{widetext}
in the basis of
\begin{eqnarray}
 \{&&\lvert r_+r_+r_+\rangle, \lvert r_+r_+\text{c}_\uparrow\rangle, \lvert r_+\text{c}_\uparrow r_+\rangle, \lvert \text{c}_\uparrow  r_+r_+\rangle,\lvert  r_+\text{c}_\uparrow \text{c}_\uparrow\rangle, \nonumber\\
 &&\lvert  \text{c}_\uparrow r_+\text{c}_\uparrow\rangle,\lvert  \text{c}_\uparrow\text{c}_\uparrow r_+\rangle ,\lvert  \text{c}_\uparrow\text{c}_\uparrow \text{c}_\uparrow\rangle \}.\nonumber
\end{eqnarray}
 For $\lvert \text{c}_\downarrow \text{c}_\downarrow\text{c}_\downarrow\rangle\equiv|\text{ccc}\rangle_{\text{e}}\otimes \lvert\downarrow\downarrow\downarrow\rangle_{\text{n}}$, the Hamiltonian is similar to Eq.~(\ref{Huuu}) with the difference that $(\Delta,\Omega_{\text{c}\uparrow})$ should be replaced by $(-\Delta,\Omega_{\text{c}\downarrow})$. For $\lvert \text{c}_\uparrow\text{c}_\uparrow \text{c}_\downarrow\rangle\equiv|\text{ccc}\rangle_{\text{e}}\otimes \lvert\uparrow\uparrow\downarrow\rangle_{\text{n}}$, the Hamiltonian is
\begin{eqnarray}
   && \left(\begin{array}{cccccccc}
  3V+\Delta& \frac{\Omega_{\text{c}\downarrow}}{2}& \frac{\Omega_{\text{c}\uparrow}}{2}& \frac{\Omega_{\text{c}\uparrow}}{2}& 0& 0& 0& 0 \\
\frac{\Omega_{\text{c}\downarrow}^\ast}{2}&  V+2\Delta& 0&  0& \frac{\Omega_{\text{c}\uparrow}}{2}&  \frac{\Omega_{\text{c}\uparrow}}{2}&   0&  0\\
\frac{\Omega_{\text{c}\uparrow}^\ast}{2}& 0&  V &  0& \frac{\Omega_{\text{c}\uparrow}}{2}&  0&  \frac{\Omega_{\text{c}\uparrow}}{2}&   0\\
\frac{\Omega_{\text{c}\uparrow}^\ast}{2}& 0& 0& V  & 0 &    \frac{\Omega_{\text{c}\downarrow}}{2}&  \frac{\Omega_{\text{c}\uparrow}}{2}&   0\\
0&  \frac{\Omega_{\text{c}\uparrow}^\ast}{2}&  \frac{\Omega_{\text{c}\uparrow}^\ast}{2}& 0& \Delta&  0&   0&  \frac{\Omega_{\text{c}\uparrow}}{2}\\
0&  \frac{\Omega_{\text{c}\uparrow}^\ast}{2}& 0&  \frac{\Omega_{\text{c}\downarrow}^\ast}{2}& 0 & \Delta&  0&  \frac{\Omega_{\text{c}\uparrow}}{2}\\
0& 0&  \frac{\Omega_{\text{c}\uparrow}^\ast}{2}&    \frac{\Omega_{\text{c}\uparrow}^\ast}{2}&0& 0& -\Delta&  \frac{\Omega_{\text{c}\downarrow}}{2}\\
0& 0& 0& 0&  \frac{\Omega_{\text{c}\uparrow}^\ast}{2}&  \frac{\Omega_{\text{c}\uparrow}^\ast}{2}& \frac{\Omega_{\text{c}\downarrow}^\ast}{2} &  0
             \end{array}\right)\nonumber\\\label{Huud}
\end{eqnarray}
in the basis of
\begin{eqnarray}
 \{&&\lvert r_+r_+r_-\rangle, \lvert r_+r_+\text{c}_\downarrow\rangle, \lvert r_+\text{c}_\uparrow r_-\rangle, \lvert \text{c}_\uparrow  r_+r_-\rangle,\lvert  r_+\text{c}_\uparrow \text{c}_\downarrow\rangle, \nonumber\\
 &&\lvert  \text{c}_\uparrow r_+\text{c}_\downarrow\rangle,\lvert  \text{c}_\uparrow\text{c}_\uparrow r_-\rangle ,\lvert  \text{c}_\uparrow\text{c}_\uparrow \text{c}_\downarrow\rangle \}.\nonumber
\end{eqnarray}
The Hamiltonian for $\lvert \text{c}_x\text{c}_y \text{c}_z\rangle$ with only one of $x,y$, and $z$ being $\downarrow$ is similar to Eq.~(\ref{Huud}) by appropriate basis state arrangement. The Hamiltonian for $\lvert \text{c}_\downarrow\text{c}_\downarrow\text{c}_\uparrow \rangle$ is similar to Eq.~(\ref{Huud}) via $\pm\Delta\rightarrow\mp \Delta$, $\Omega_{\text{c}\uparrow}\rightarrow \Omega_{\text{c}\downarrow}$, and $\Omega_{\text{c}\downarrow}\rightarrow \Omega_{\text{c}\uparrow}$, and the Hamiltonian for $\lvert \text{c}_x\text{c}_y \text{c}_z\rangle$ with only one of $x,y$, and $z$ being $\uparrow$ is similar to the Hamiltonian of $\lvert \text{c}_\downarrow\text{c}_\downarrow\text{c}_\uparrow \rangle$.

For the input state $\lvert \text{c}_\uparrow\text{c}_\uparrow \text{c}_\uparrow\rangle$ which is $\frac{e^{i\alpha^{(\blacktriangle)'}}}{\sqrt{3}} (\lvert r_+ \text{c}_\uparrow\text{c}_\uparrow\rangle+ \lvert   \text{c}_\uparrow r_+\text{c}_\uparrow\rangle+\lvert   \text{c}_\uparrow\text{c}_\uparrow r_+\rangle)$ at the beginning of the third pulse, the Hamiltonian is a $6\times6$ matrix written in the basis of $ \{ \lvert r_+\text{c}_\uparrow\text{c}_\uparrow\rangle, \lvert \text{c}_\uparrow r_+\text{c}_\uparrow \rangle,  \lvert \text{c}_\uparrow\text{c}_\uparrow r_+\rangle, \lvert \text{g}_\uparrow \text{c}_\uparrow \text{c}_\uparrow\rangle, \lvert \text{c}_\uparrow \text{g}_\uparrow \text{c}_\uparrow\rangle , \lvert \text{c}_\uparrow  \text{c}_\uparrow\rangle\text{g}_\uparrow \} $
in which three pairs of states are coupled in a way similar to that of Eq.~(\ref{H-third-uu}). 
For the input state $\lvert \text{c}_\downarrow \text{c}_\downarrow \text{c}_\downarrow\rangle$ which is  $\frac{ e^{-i\alpha^{(\blacktriangle)'}}   }{\sqrt{3}} (\lvert r_- \text{c}_\downarrow\text{c}_\downarrow\rangle + \lvert \text{c}_\downarrow r_- \text{c}_\downarrow\rangle + \lvert   \text{c}_\downarrow\text{c}_\downarrow r_- \rangle )$ at the beginning of the third pulse, the Hamiltonian is written in a way similar to that in Eq.~(\ref{H-third-dd}) in an appropriate six-ket basis.

\begin{figure}
\includegraphics[width=3.2in]
{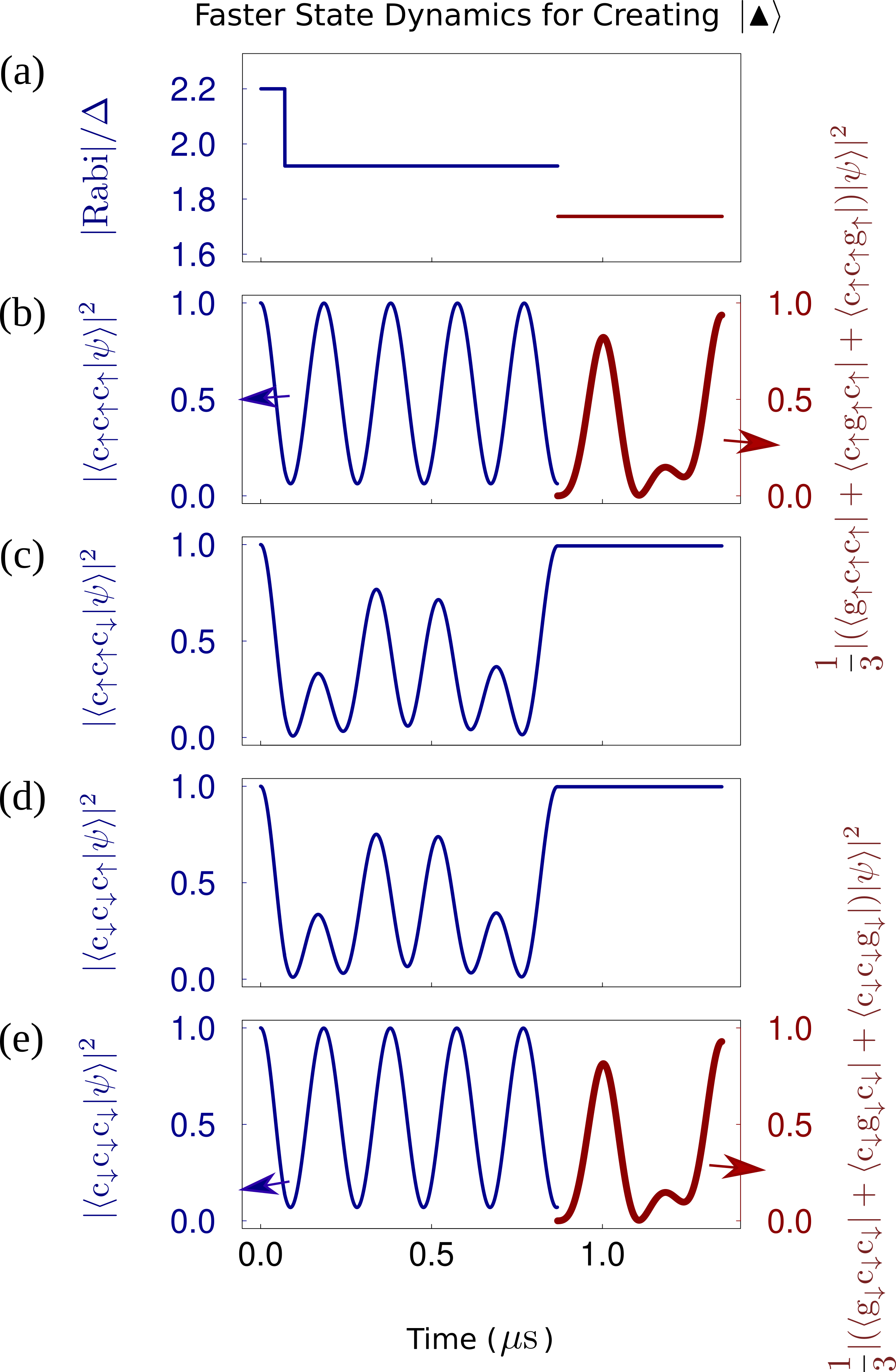}
 \caption{State dynamics of the $\lvert\blacktriangle\rangle$ protocol with a faster speed with $V_0/2\pi=260$~MHz and maximal Rabi frequency $\Omega^{(\blacktriangle)}/2\pi\approx3.25$~MHz when ignoring Rydberg-state decay. The meaning of the contents are the same as those in Fig.~\ref{W-unitary}. In (b) and (e), there is some population not transferred to the Rydberg state at the end of the second pulse, which results in a larger error of the entanglement generation compared to the case described around Fig.~\ref{W-unitary}.   }\label{W-unitary-F}
\end{figure}

\section{A faster protocol for creating $\lvert\blacktriangle\rangle$}\label{AppendixE}
\begin{figure}
\includegraphics[width=2.5in]
{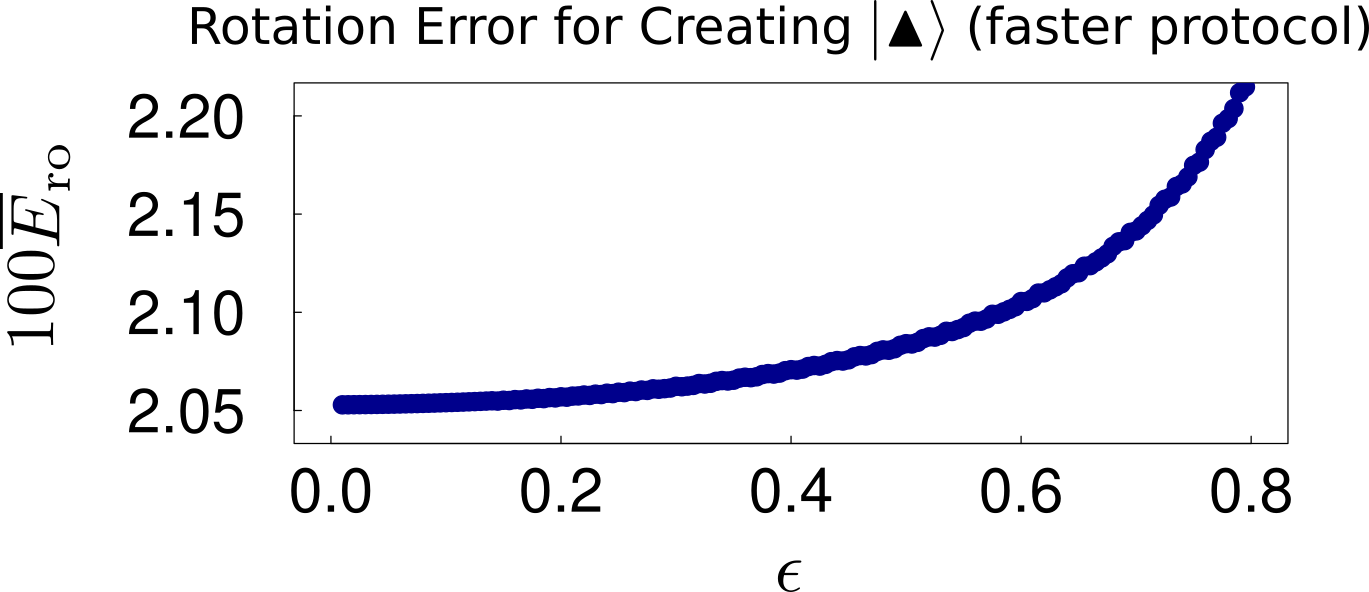}
 \caption{ Rotation error~(rescaled by $100$) averaged by uniformly varying the Rydberg interaction $V$ in $[(1-\epsilon)V_0,~(1+\epsilon)V_0]$ for creating $\lvert\blacktriangle\rangle$ with the parameters of Fig.~\ref{W-unitary-F} and Appendix~\ref{AppendixD}. The fidelity to generate $\lvert\blacktriangle\rangle$ in the form similar to that in Eq.~(\ref{W-final-real}), $1- \overline{E}_{\text{ro}} -  E_{\text{decay}} $, is about $0.976$ for $\epsilon=0.8$. }\label{W-rotation-Error-Appendix}
\end{figure}

The protocols in Secs.~\ref{Sec02},~\ref{Sec03}, and~\ref{Sec04} describe high-fidelity methods. Sometimes a fast entangling gate is helpful for certain purposes. One approach to faster entanglement is to find approximate excitation of Rydberg states. We take the generation of $\lvert\blacktriangle\rangle$ as an example. In Sec.~\ref{Sec04}, the first pulse has a long duration so as to have a high-fidelity generation of the Rydberg states in Eq.~(\ref{W-pulse21}). One can use a shorter pulse to achieve it though the fidelity is a little lower. We find that with the three Rabi frequencies equal to $(2.2, 1.92, 1.737)\Delta$ and the three pulse durations $(0.206, 2.36, 1.421)\pi/\Delta$, we can approximately realize the state $\lvert \blacktriangle\rangle$ with a total duration $8.8\pi/\Omega_{\text{m}}$, where $\Omega_{\text{m}}$ is the maximal $\Omega$~(of the first pulse). With these parameters and the same maximal Rabi frequency used in Fig.~\ref{W-unitary}, we simulate the state dynamics, and found that the final populations are $0.937$, $0.993$, $0.997$, and $0.929$ respectively for the four typical types of input states $\lvert\text{c}_\uparrow \text{c}_\uparrow\text{c}_\uparrow  \rangle, \lvert\text{c}_\uparrow \text{c}_\uparrow\text{c}_\downarrow  \rangle,\lvert\text{c}_\downarrow \text{c}_\downarrow\text{c}_\uparrow  \rangle,\lvert\text{c}_\downarrow \text{c}_\downarrow\text{c}_\downarrow  \rangle$, shown in Fig.~\ref{W-unitary-F}. The Rydberg-state decay induces an error $E_{\text{decay}}\approx1.79\pi/(\tau\Delta)$, which is about $2.0\times10^{-3}$ when $\Omega_{\text{eff}}/2\pi=1$~MHz and $\tau=330~\mu$s. For this case, the main error to the entanglement generation is the population loss which can be obviously seen in Figs.~\ref{W-unitary-F}(b) and~\ref{W-unitary-F}(d), which leads to large rotation errors as shown in Fig.~\ref{W-rotation-Error-Appendix}. With a large interaction fluctuation $\epsilon=0.8$, the fidelity is about $0.976$. We would like to point out that by pulse shaping techniques~\cite{Shi2021qst} one can in principle design much faster protocols with high fidelity, but the purpose of the second part of this paper is to show that the near degeneracy of the nuclear spin qubits in a weak magnetic field can enable fast generation of exotic entanglement between nuclear spins and electrons in multiple atoms.

%


\end{document}